\documentclass[10pt,letterpaper]{article}
\usepackage[top=0.85in,left=2.75in,footskip=0.75in]{geometry}
\usepackage{floatrow}
\usepackage{amsmath,amssymb}

\usepackage{changepage}

\usepackage[utf8x]{inputenc}

\usepackage{textcomp,marvosym}

\usepackage{cite}

\usepackage{nameref,hyperref}

\usepackage[right]{lineno}

\usepackage{microtype}
\DisableLigatures[f]{encoding = *, family = * }

\usepackage[table]{xcolor}

\usepackage{array}

\newcolumntype{+}{!{\vrule width 2pt}}

\newlength\savedwidth

\raggedright
\usepackage[aboveskip=1pt,labelfont=bf,labelsep=period,justification=raggedright,singlelinecheck=off]{caption}

\makeatletter
\renewcommand{\@biblabel}[1]{\quad#1.}
\makeatother

\date{}

\usepackage{lastpage,fancyhdr,graphicx}
\usepackage{epstopdf}
\fancyheadoffset[L]{2.25in}
\fancyfootoffset[L]{2.25in}
\newcommand{\nwc}{\newcommand}
\nwc{\av}[1]{\left< #1\right>}

\newcommand{\be}{\begin{equation}}
\newcommand{\ee}{\end{equation}}
\newcommand{\nn}{\nonumber}
\newcommand{\ba}{\begin{align}}
\newcommand{\ea}{\end{align}}

\begin{document}
\vspace*{0.2in}

\begin{flushleft}
{\Large
\textbf\newline{Information-theoretic analysis of the directional influence between  cellular processes} 
}
\newline
\\
Sourabh Lahiri \textsuperscript{1},
Philippe Nghe\textsuperscript{2},
Sander J. Tans\textsuperscript{3},
Martin Luc Rosinberg\textsuperscript{4},
David Lacoste\textsuperscript{1*},
\\
\bigskip
\textbf{1} Gulliver laboratory, PSL Research University, ESPCI, 10 rue de Vauquelin, 75231 Paris Cedex 05, France
\\
\textbf{2} Laboratory of Biochemistry, PSL Research University, ESPCI, 10 rue de Vauquelin, 75231 Paris Cedex 05, France
\\
\textbf{3} FOM Institute AMOLF, Science Park,104, 1098 XG Amsterdam, the Netherlands
\\
\textbf{4} Laboratoire de Physique Th\'eorique de la Mati\`ere Condens\'ee,
Universit\'e Pierre et Marie Curie, CNRS UMR 7600, 4 place Jussieu, 75252 Paris Cedex 05, France
\\
\bigskip

%
%





* david.lacoste@espci.fr

\end{flushleft}
\section*{Abstract}
Inferring the directionality of interactions between cellular processes is a major challenge in systems biology. 
Time-lagged correlations allow to discriminate between alternative models, but they still rely on assumed underlying interactions. 
Here, we use the transfer entropy (TE), an information-theoretic quantity that quantifies the directional influence 
between fluctuating variables in a model-free way. We present a theoretical approach to compute the transfer entropy, even when the noise has an extrinsic component or in the presence of feedback. We re-analyze the experimental data from Kiviet et al. (2014) where fluctuations in gene expression of metabolic enzymes and growth rate have been measured in single cells of {\it E. coli}. We confirm the formerly detected modes between growth and gene expression, while prescribing more stringent conditions on the structure of noise sources. We furthermore point out practical requirements in terms of length of time series and sampling time which must be satisfied in order to infer optimally transfer entropy from times series of fluctuations.



\section*{Introduction}
Quantifying information exchange between
variables is a general goal in many studies of biological systems because the complexity of such systems
prohibits mechanistic bottom-up approaches.
Several statistical methods have been proposed to exploit either the specific dependence of the covariances 
between input and output variables with respect to a perturbation applied to the network \cite{Prill2015}, or 
the information contained in 3-point correlations \cite{Affeldt2016}. These methods are potentially well suited for datasets 
obtained from destructive measurements, such as RNA sequencing or immunohistochemistry. 

However, none of these methods exploits the information contained in time-lagged statistics, which is provided for instance by non-destructive 
measurements obtained from time-lapse microscopy of single cells. Such experimental data should be quite relevant to understand functional relationships 
since they merely reflect the time delays present in the dynamics of the system. Time-delayed cross-correlations between gene expression fluctuations 
have indeed been shown to discriminate between several mechanistic models of well characterized genetic networks~\cite{Dunlop2008}. 
However, such methods become difficult to interpret in the presence of feedback. 

This situation is illustrated in reference~\cite{kiv14_nature} where the fluctuations in the growth rate and in the expression level of metabolic 
enzymes have been measured as a function of time by tracking single cells of {\it E. coli} 
with time-lapse microscopy. The interplay between these variables has been characterized using cross-correlations as proposed in \cite{Dunlop2008}. 
To circumvent the difficulty of discriminating between many complex and poorly parametrized metabolic models, the authors reduced functional 
relations to effective linear responses with a postulated form of effective couplings. 

In the present work, we instead use a time-lagged and information-based method to analyze the interplay between the two 
fluctuating variables. A crucial feature in this method is that it is model-free 
and it is  
 able to disentangle the two directions of influence between the two variables,  unlike the cross-correlations discussed above. 
This type of approach was first proposed by Granger~\cite{Granger1969} in the field of econometrics and found applications in a broader area. 
More recently, transfer entropy~\cite{sch00_prl}, which is a non-linear extension of Granger causality, 
has become a popular information-theoretic measure to infer directional relationships between jointly dependent processes~\cite{wib13_plosone}. 
It has been successfully applied to various biomedical time series (see for instance \cite{Pahle2008}) and used 
 extensively in the field of neurobiology, as shown in Ref. \cite{Vicente2011} and in references therein. 
This is the tool that will be used in this work. 

The plan of this paper is as follows. We first introduce two measures of information dynamics, transfer entropy (TE) and information flow (IF).  
We then illustrate our numerical method on a well controlled case, namely a simple linear Langevin model, and show that we can properly 
estimate these quantities from the generated time series.
We then analyze experimental data on the fluctuations of 
metabolism of {\it E. coli} taken from Ref.~\cite{kiv14_nature}.
We provide analytical expressions for the transfer entropy and 
information flow rates for the model proposed in that reference. 
After identifying a divergence in one TE rate as the sampling time goes to zero, 
we introduce a simplified model which 
is free of divergences while still being compatible with the experimental data. 
We conclude that the inference of information-theoretic dynamical quantities can 
be helpful to build physically sound models of the various noise components present in chemical networks.

\subsection*{Information theoretic measures}

Unlike the mutual information $I(X:Y)$ that only quantifies the amount of  information exchanged between two random variables $X$ and $Y$ as 
defined in the section on Methods, the transfer entropy (TE) is an asymmetric measure that can discriminate between a source and a target \cite{sch00_prl}. 
Consider two sampled time series $\{..x_{i-1},x_i,x_{i+1}..\}$ and $\{..y_{i-1},y_i,y_{i+1}..\}$, where $i$ is the discrete time index, generated by a source process $X$ and  a target process $Y$. 
The transfer entropy $T_{X \to Y}$ from  $X$ to $Y$ is a conditional, {\it history-dependent} mutual information defined as
%
\begin{align}
 T_{X \to Y}&=\sum P(y_{i+1},\pmb{y}_i^{(k)},\pmb{x}_i^{(l)})
\ln\frac{P(y_{i+1}|\pmb{y}_i^{(k)},\pmb{x}_i^{(l)})}{P(y_{i+1}|\pmb{y}_i^{(k)})}, \nonumber \\
&=\sum_i \: [ H(y_{i+1}|\pmb{y}_i^{(k)})-H(y_{i+1}|\pmb{y}_i^{(k)},\pmb{x}_i^{(l)})]
\label{def TE}
\end{align}
where $\pmb{y}_i^{(k)}=\{y_{i-k+1},\cdots, y_i\}$ and $\pmb{x}_i^{(l)}=\{x_{i-l+1},\cdots, x_i\}$ denote two blocks  of past values of 
$Y$ and $X$ of length $k$ and $l$ respectively, $P(y_{i+1},\pmb{y}_i^{(k)},\pmb{x}_i^{(l)})$ is the joint probability of observing 
$y_{i+1}, \pmb{y}_i^{(k)},\pmb{x}_i^{(l)}$, and $P(y_{i+1}|\pmb{y}_i^{(k)},\pmb{x}_i^{(l)}), P(y_{i+1}|\pmb{y}_i^{(k)})$ are conditional  probabilities. 
In the second line, $H(.\vert .)$ denotes the conditional Shannon entropy (see Section on Methods for definition).
In the first equation, the summation is taken over all possible values of the random variables $y_{i+1},\pmb{y}_i^{(k)},\pmb{x}_i^{(l)}$ and over all values of the time index $i$.

To put it in simple terms, $T_{X\to Y}$ quantifies the {\it information contained from the past of $X$ about the 
future of $Y$, which the past of $Y$ did not already provide} \cite{wib13_plosone,Pahle2008}. Therefore, it should  be regarded as a measure of {\it predictability} 
rather than a measure of {\it causality} between two time-series \cite{Lizier2010}.
For instance, when $\pmb{x}_i^{(l)}$  does not bring new information on $y_{i+1}$, then 
$P(y_{i+1}|\pmb{y}_i^{(k)},\pmb{x}_i^{(l)})=P(y_{i+1}|\pmb{y}_i^{(k)})$ and the transfer entropy vanishes because  the prediction on $y_{i+1}$ is not improved.
With a similar definition for $T_{Y\to X}$, one can define
the net variation of transfer entropy from $X$ to $Y$ as
$\Delta T_{X\to Y}\equiv T_{X\to Y}-T_{Y\to X}$. The sign of $\Delta T_{X\to Y}$
informs on the directionality of the information transfer.

The statistics required for properly evaluating the transfer entropy rapidly increases with $k$ and $l$, 
which in practice prohibits the use of large values of $k$ and $l$. The most accessible case  thus corresponds to $k=l=1$, which we denote 
hereafter as $\overline T_{X\to Y}$. This quantity is then simply defined as 
\begin{align}
 \overline T_{X\to Y}=\sum_i \big [H(y_{i+1} | y_i)- H(y_{i+1} | y_i,x_i)\big],
\end{align}
When the dynamics of the joint process $\{ X,Y \}$ is Markovian, one has 
$P(y_{i+1}|\pmb{y}_i^{(k)},\pmb{x}_i^{(l)})=P(y_{i+1}| y_i,x_i)$ and since $H(y_{i+1}|\pmb{y}_i^{(k)}) \le 
H(y_{i+1}| y_i)$ one has $\overline T_{X\to Y} \ge T_{X\to Y}$ 
(see Ref. \cite{Hartich2016}). 
Therefore, $\overline T_{X\to Y}$ represents an upper bound 
on the transfer entropy. In the case of stationary time series, which is the regime we consider in this work, it is natural to also introduce the TE rate
\begin{align}
{\overline {\cal T}}_{X\to Y}&= \lim_{\tau \to 0} \frac{H(y_{t+\tau} | y_t)- H(y_{t+\tau} | x_t,y_t) }{\tau}\nonumber\\
&=\lim_{\tau \to 0} \frac{I(y_{t+\tau}:y_t,x_t)- I(y_{t+\tau}:y_t) }{\tau}\ ,
 \label{TE rate}
\end{align}
where the continuous time variable $t$ replaces the discrete index $i$. In practice 
${\overline {\cal T}}_{X\to Y} \simeq {\overline T}_{X\to Y}/\tau$, 
but only for sufficiently small time step $\tau$.

The most direct strategy to evaluate Eq. (\ref{def TE}) would be to 
construct empirical estimators of the probabilities from histograms of the data. 
Although this procedure works well for evaluating other quantities, for instance the entropy production in small 
stochastic systems~\cite{kun14_prl}, it completely fails in the case of transfer entropy. Indeed, such a method  leads to    
a non-zero TE even between uncorrelated signals, due to strong biases in standard estimators based on data binning.
In order to overcome this problem, we used the Kraskov-St\"ogbauer-Grassberger (KSG) 
estimator which does not rely on binning, as implemented in the  software package JIDT (Java Information Dynamics Toolkit) \cite{jidt}.
Using estimators of this kind is particularly important for variables that take continuous values.

In the following, the inference method will be applied to time series generated by diffusion processes. It will then be interesting to compare the TE rate ${\overline {\cal T}}_{X\to Y}$ to another measure of information dynamics, the so-called information flow~\cite{Parrondo2015,Horowitz2014a,hor14_njp} (also dubbed learning rate in the context of sensory systems~\cite{Hartich2014,Hartich2016}), which is defined as the time-shifted mutual information~\cite{Allahverdyan2009}
\be
{\cal I}^{flow}_{X \to Y}=\lim_{\tau \to 0} \frac{I(y_t:x_t)- I(y_t:x_{t+\tau}) }{\tau}\ .
\ee
In the special case where the two  processes $X$ and $Y$ experience independent noises (the system is then called {\it bipartite})~\cite{Horowitz2014a}, one has the inequality ${\cal I}^{flow}_{X \to Y} \le {\cal T}_{X \to Y}$~\cite{Hartich2014}, which in turn implies  that
\be
{\cal I}^{flow}_{X \to Y} \le \overline {\cal T}_{X \to Y} 
\label{ineq Iflow}
\ee 
when the joint process is Markovian. Observing a violation of this inequality is thus a strong indication that the noises on $X$ and $Y$ are correlated. As will be seen later, this is indeed the situation in biochemical networks, due the presence of the so-called extrinsic noise generated  by the stochasticity in the cell and in the cell environment~\cite{Bowsher2012} which acts on all chemical reactions within the cell, and thus induces correlations.


\section*{Results}
\subsection*{Test of the inference method on a Langevin model}

In order to benchmark our inference method and perform a rigorous test in a controlled setting, we first applied it on times series generated by a simple model for which the transfer entropy and the information flow can be computed analytically. The data were obtained by simulating the two coupled Langevin equations
\begin{align}
\label{Lang}
m\dot v &= -\gamma v-a y+\xi, \nn \\
\tau_r\dot y &=  v-y+\eta
\end{align}
that describe the dynamics of a particle of mass $m$ subjected to a velocity-dependent feedback that damps thermal fluctuation 
\cite{Kim2004,Munakata2013,hor14_njp} (in these equations, the dependence of the variables on the time $t$ is implicit). Here, $\xi(t)$ is the noise generated 
by the thermal environment with viscous damping $\gamma$ and temperature $T$, while $\eta(t)$ is the noise associated with the measurement of the 
particle's velocity $v(t)$. The two noises are independent and Gaussian with zero-mean and variances $\av{\xi(t)\xi(t')}=2\gamma k_BT\delta(t-t')$ and 
$\av{\eta(t)\eta(t')}=\sigma^2\delta(t-t')$. $a$ is the feedback gain and $\tau_r$ is a time constant.
\begin{figure}[!h]
  \centering
 {\includegraphics[width=7cm]{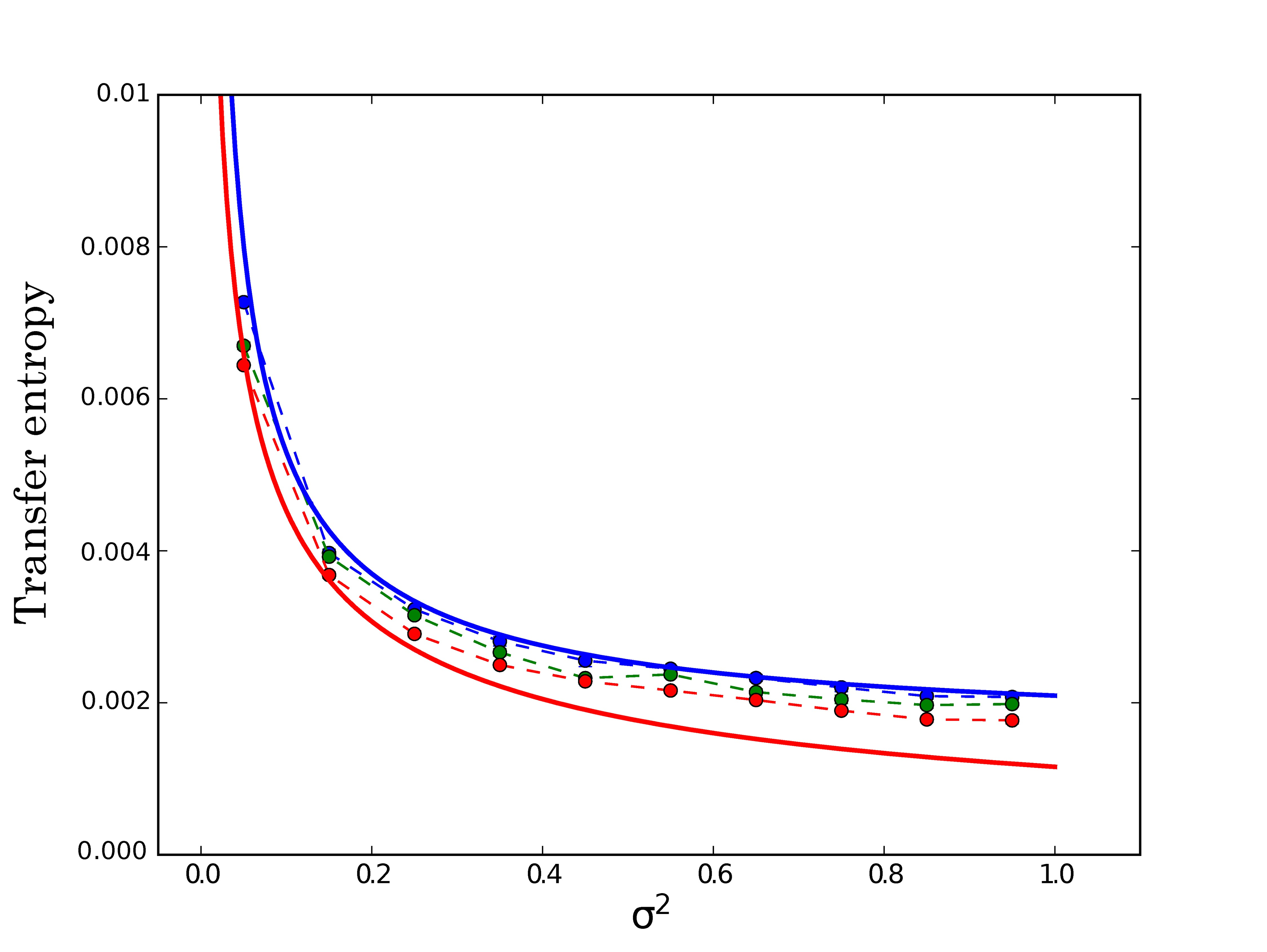}}
  \caption{Transfer entropy $T_{Y\to V}$ for the feedback model governed by Eqs. (\ref{Lang}) as a function of 
the noise intensity  $\sigma^2$ for $k=1$ (blue circles), $k=3$ (green circles) and $k=5$ (red circles). The parameter $l$ 
present in the definition of Eq. (\ref{def TE}) is fixed to 1.
The lower red (resp. upper blue) solid line represents the value of $T_{Y\to V}$  (resp. $\overline T_{Y\to V}$) obtained by multiplying the theoretical rate ${\cal T}_{Y\to V}$ (resp. $\overline {\cal  T}_{Y\to V}$) given by Eq. (\ref{TEfeedback}) (resp. Eq. (\ref{TEbar}) by the sampling time $\tau=10^{-3}$. 
 The parameters of the model are 
 $T=5$, $\gamma=m=1$, $\tau_r=0.1$, and $a=8$. 
 \label{fig:te_a_jidt} }
\end{figure}

The two Langevin equations were numerically integrated with the standard Heun's method \cite{Sauer2013} using a time step $\Delta t=10^{-3}$, and the transfer entropy in the steady state was estimated from $100$
 time series of duration $t=2000$ with a sampling time (i.e., the time between two consecutive data points) $\tau= \Delta t$. We first checked that the TE  in the direction $Y \to V$ does vanish in the absence of feedback, i.e. for $a=0$, whereas it is non-zero as soon as $a>0$. We then tested the influence of the measurement error $\sigma^2$ for a fixed value of the gain $a$. 
As can be seen in Fig \ref{fig:te_a_jidt},  $T_{V \to Y}$  diverges as $\sigma^2 \to 0$, a feature that will play an important role in our discussion of the model for the metabolic network.
In the figure, the color of the symbols correspond to three different values of the parameter $k$ which represents the history length in the definition of the transfer entropy (see Eq. (\ref{def TE})).
One can see that the estimates of $T_{V \to Y}$ for $k=1$ are in very good agreement with the theoretical prediction for 
$\overline T_{V \to Y}$ (upper solid line). Moreover, the estimates decrease as $k$ is increased from $1$ to $5$, and one
 can reasonably expect that the theoretical value of  $T_{V \to Y}$ (lower solid line) computed in Ref.~\cite{hor14_njp} and given by
 Eq.~(\ref{TEfeedback}) in the section on Methods would be reached in the limit $k \to \infty$.

Finally, by estimating the information flow and the transfer entropy, we checked that inequality (\ref{ineq Iflow}) 
holds, as a result of the independence of the 
two noises $\xi$ and $\eta$ (see section on Methods).

\subsection*{Analysis of stochasticity in a metabolic network}
\subsubsection*{Experimental time series}
We are now in position to analyze the fluctuations in the metabolism of {\it E. coli} 
at the single cell level obtained in Ref. \cite{kiv14_nature} using the information-theoretic 
notions introduced and tested in the previous section. 
Since there are a multitude of reactions and interactions involved in the metabolism of {\it E. coli}, a complete mechanistic 
description is not feasible, and our model-free inference method has a crucial advantage.
In Ref. \cite{kiv14_nature},  the length of the cells was recorded as a function of time using image analysis, 
and the growth rate was then obtained by fitting this data over subparts of the cell cycle. In the same experiment, the fluorescence level of 
GFP, which is co-expressed with growth enzymes LacY and LacZ 
was recorded.
Three set of experiments were carried out corresponding to three levels of an inducer IPTG: low, intermediate and high. 
\begin{figure}[!h]
  \centering
  {\rotatebox{-0}{\includegraphics[width=7cm]{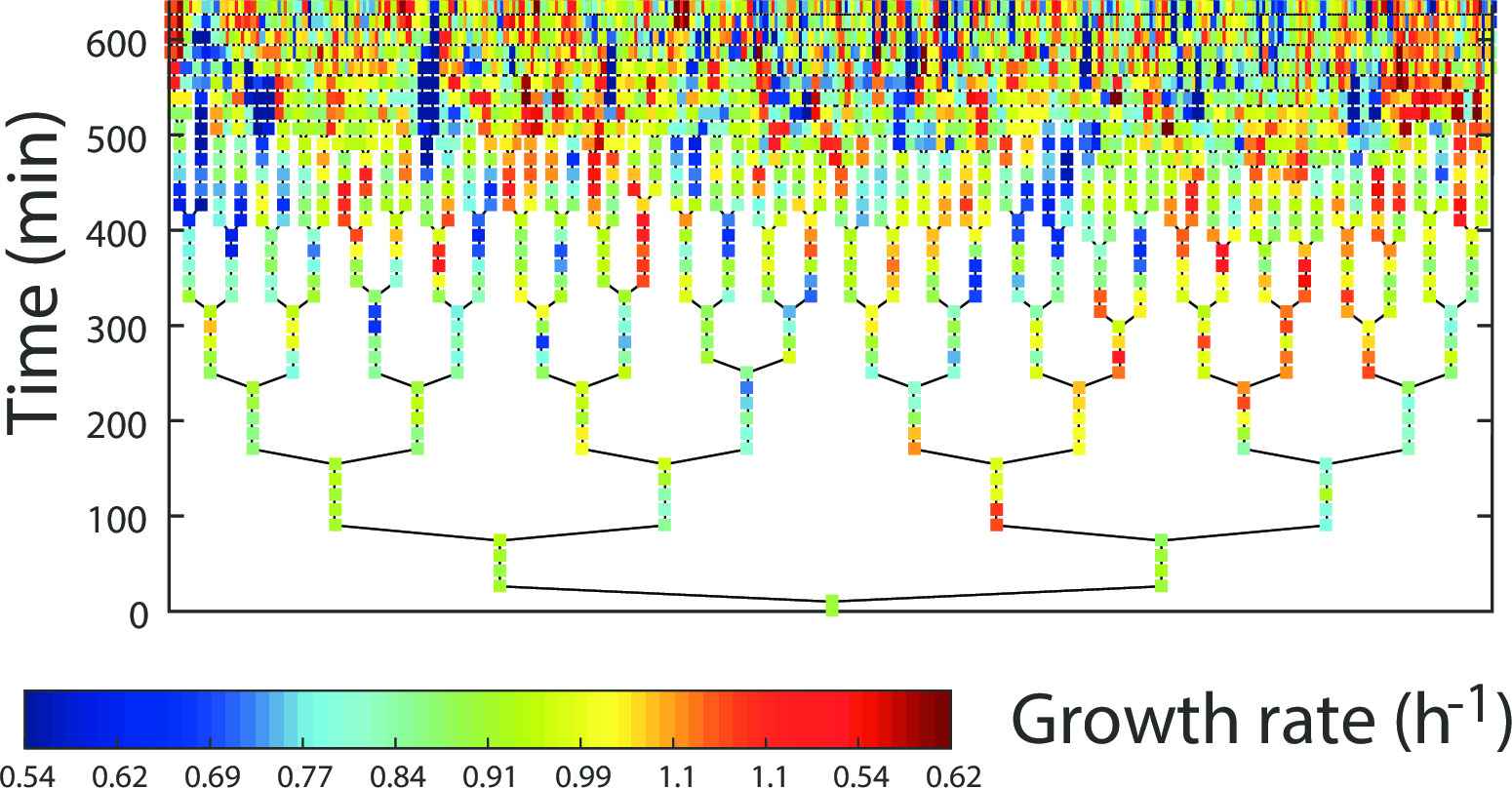}}}
  \caption{Pedigree tree representing the evolution of the colony of {\it E. coli}. studied in Ref. \cite{kiv14_nature}. The splitting 
of the branches corresponds to cell division events, each colored point is associated to a measurement of a single cell 
and the colors represent the growth rates as shown in the bar in the lower part of the figure.
  \label{fig:tree}
  }
\end{figure}

The two time series have a branching structure due to the various lineages, which all start from a single mother cell as shown in Fig \ref{fig:tree}.
The experimental data thus come in the form of a large ensemble of short times series which represent a record of all the cell cycles. 
There are about $\sim 3000$ time series, with 2 to 8 measurement points in each of them which are represented as colored points in Fig \ref{fig:tree}. 
In order to correctly estimate the transfer entropy from such data, we have analyzed the multiple time series as 
 independent realizations of the same underlying stochastic process. 
For the present analysis, we fix the history length parameters 
$k$ and $l$ to the value $k=l=1$, which means that we focus on $\overline T$ rather than $T$. 
We infer the values of $\overline T$ in the two directions, from growth (denoted $\mu$) to gene expression (denoted $E$) and vice versa.
The results obtained for the three concentrations of IPTG 
are represented in Table \ref{table:unconcat}. The negative value of $\overline T_{\mu\to E}$ which is found in the intermediate case is due to the numerical inference method and should be regarded as a value which cannot be distinguished from zero.

\begin{center}
\begin{table}[!h]
\renewcommand{\arraystretch}{1.5}
\begin{tabular}{|c|c|c|c|}
 \hline
  {\bf Conc. of IPTG} & {Low} & {Intermediate} & {High}\\
  \hline
  $\overline T_{E\to\mu}$ & $2.35\cdot 10^{-2}$ & $1.37\cdot 10^{-2}$& $1.06\cdot 10^{-3}$\\
  \hline
  $\overline T_{\mu\to E}$ & $2.16\cdot 10^{-2}$& $-4.08\cdot 10^{-3}$& $9.94\cdot 10^{-3}$\\
  \hline
  $\Delta \overline T_{E\to\mu}$ & $1.84\cdot 10^{-4}$& $1.78\cdot 10^{-2}$& $-8.88\cdot 10^{-3}$ \\
  \hline
\end{tabular}
\caption{Inferred values of the transfer entropies in the directions $E\to \mu$ and 
$\mu\to E$, and the difference $\Delta \overline T_{E\to\mu}=\overline T_{E\to\mu}-\overline T_{\mu\to E}$ 
for low, medium and high concentrations of IPTG based on the data of ref. \cite{kiv14_nature}. The TE are given in nats.}
\label{table:unconcat}
\end{table}
\end{center}

Based on this analysis, we conclude that the influence between the variables is 
directed primarily from enzyme expression to growth 
in the low and intermediate IPTG experiments, 
while it mainly proceeds in the reverse direction in the high IPTG experiment. 
Such results are in line with the conclusions of Ref. \cite{kiv14_nature} based on the measured asymmetry of the time-lagged cross-correlations. Moreover, the present analysis provides an estimate of the influence between the two variables
separately in the two directions from $E$ to $\mu$ and from $\mu$ to $E$.   
In particular, we observe for the low experiment that the values of TE in the two directions are of same order of magnitude, 
whereas in the intermediate experiment the TE from $E$ to $\mu$ is larger, 
a feature which could not have been guessed from measured time delays.



\subsubsection*{Theoretical Models}
We now turn to the analysis of the model proposed in Ref.~\cite{kiv14_nature} to account for the experimental data. 
The question we ask is whether the model correctly reproduces the above results for the transfer entropies, in particular the change 
in the sign of $\Delta \overline T_{E\to \mu}$ for the  high concentration of IPTG. 

The central equation of the model describes the production of the enzyme as
\begin{align}
  \dot E = p-\mu\cdot E,
  \label{eq:Edot}
\end{align}
where $E$ is the enzyme concentration, $p$ its production rate, and $\mu$ the rate of increase in cell volume. Although the function $p$ is typically 
non-linear, its precise expression is irrelevant because \eqref{eq:Edot} is linearized around the stationary point defined by the mean values 
$E=E_0$ and $\mu=\mu_0$. 
This linearization then yields 
\begin{align}
\label{linearized}
\delta \dot E = \delta p - \delta \mu E_0 - \mu_0 \delta E, \, 
\end{align} 
in terms of perturbed variables $\delta X(t)=X(t)-X_0$, where $X_0$ denotes the mean of $X$.

The model of Ref.~\cite{kiv14_nature} is essentially phenomenological in nature because it approximates the noises as Gaussian processes. 
Although this approximation is often done in this field, it may not always hold since fluctuations due to 
low copy numbers are generally not Gaussian \cite{Monteoliva2017}.
In any case, the model contains three Gaussian noises: $N_G$ is a common component while
$N_E$ and $N_\mu$ are component specific to $E$ and $\mu$. These noises are assumed to be 
independent  Ornstein-Uhlenbeck noises with  zero mean and autocorrelation 
functions $\langle N_i(t)N_i(t')\rangle=\eta_i^2e^{-\beta_i\vert t-t'\vert}$ ($i=E,\mu,G$).
As commonly done, the three Ornstein-Uhlenbeck noises are generated by the auxiliary equations 
\begin{align}
\label{EqOU}
\dot N_i=-\beta_i N_i+\xi_i, \, 
\end{align} 
where the $\xi_i's$ are  zero-mean Gaussian white noises satisfying $\langle \xi_i(t)\xi_j(t')\rangle=\theta_i^2\delta(t-t')\delta_{ij}$ with 
$\theta_i=\eta_i\sqrt{2\beta_i}$.  
Introducing the constant logarithmic gains $T_{XY}$ that represent how a variable $X$ responds 
to the fluctuations of a source $Y$, the equations of the model read \cite{kiv14_nature} 
\begin{align}
\label{postulated}
\frac{\delta p}{E_0 \mu_0} &= T_{EE} \frac{\delta E}{E_0} + T_{E G} N_G + N_E, \nonumber \\
\frac{\delta \mu}{\mu_0} &= T_{\mu E} \frac{\delta E}{E_0} + T_{\mu G} N_G + N_\mu, 
\end{align}
where specifically $T_{E\mu}=-1$ and $T_{\mu G}=1$. 
Then, eliminating $\delta p$ from Eqs.~\eqref{linearized} and \eqref{postulated}, one obtains
 the coupled equations 
\begin{align}
\label{Eqmodel} 
\dot x&=\mu_0\big[(T_{EE}-1)x+T_{E\mu}y+T_{EG}N_G+N_E\big]\nonumber\\
y&=T_{\mu E}x+T_{\mu G} N_G+N_{\mu},
\end{align}
where we have defined 
the reduced variables $x=\delta E/E_0$, $y=\delta \mu /\mu_0$.
We stress that  $N_G$ is an {\it extrinsic} noise that affects both the enzyme concentration and the growth rate, 
whereas $N_E$ (resp. $N_{\mu}$) is an {\it intrinsic} noise that only affects $E$ (resp. $\mu$). 
Note that the two effective noises $T_{EG}N_G+N_E$ and $T_{\mu G}N_G+N_{\mu}$ acting on $\dot x$ and $y$ 
are colored {\it and} correlated, which makes the present model more complicated than most stochastic models studied in the current literature.
In fact, since we are mainly interested in the information exchanged between  $x$ and $y$, it is convenient to replace one of the noises, 
say $N_G$, by the dynamical variable $y$. 
Differentiating the second equation in Eq.~\eqref{Eqmodel}, using Eq.~\eqref{EqOU} and performing 
some simple manipulations, one then obtains a new set of equations for the four
random variables $x,y, u\equiv N_E,v\equiv N_{\mu}$:
\begin{align}
\label{Eqxuvy}
\dot x&=a_1x+a_2 u +a_3v + a_4y\nonumber\\
\dot y&=b_1x+b_2 u +b_3v + b_4y+\xi_{y}\nonumber\\
\dot u&=-\beta_E u+\xi_E\nonumber\\
\dot v&=-\beta_{\mu} v+\xi_{\mu}\ ,
\end{align}
where the coefficients $a_j$ and $b_j$ ($j=1...4$) are defined by Eqs. (\ref{coeff a_i}) in the section on Methods
and $\xi_{y}=\xi_{\mu}+\xi_G$ is a new white noise satisfying $\langle \xi_{y}(t)\xi_{y}(t')\rangle=(\theta_{\mu}^2+\theta_G^2)\delta(t-t')$ and 
$\langle \xi_{y}(t)\xi_{\mu}(t')\rangle=\theta_{\mu}^2\delta(t-t')$.  

The calculation of the transfer entropy rate ${\overline {\cal T}}_{X\to Y}$ (which coincides with ${\overline {\cal T}}_{E\to \mu}$ 
since the TE is invariant under the change of variables from $E$ to $x$ and $\mu$ to $y$) is detailed in the section on Methods, 
together with the calculation of the information flows. The final expression reads
\begin{align} 
\label{Eqfinal1}
{\overline {\cal T}}_{X\to Y}&=\frac{1}{4(\beta_{\mu}\eta_{\mu}^2+\beta_G \eta_G^2)}\int dx\:dy\: p(x,y)\big[\bar g_{y}^2(x,y)-{\bar {\bar g}}_{y}^2(y)\big]\,
\end{align}
where $p(x,y)$ is the steady state probability distribution and the functions $\bar g_{y}$ and ${\bar {\bar g}}_{y}$ are defined in Eqs. (\ref{g bar}) and 
(\ref{g bar bar}), respectively. This result agrees with that obtained in Refs.~\cite{Hartich2016}, \cite{Allahverdyan2009} and in \cite{ito14_nc} in special cases.

In Table \ref{table OU}, we show the results of the analysis of the time series generated by Eqs. (\ref{Eqxuvy})
using our numerical inference method with a sampling time $\tau=1$min (equal to the time step $\Delta t$ used to numerically integrate the model).  One can see that the estimates  of ${\overline {\cal T}}_{E\to \mu}$  are in good agreement with the predictions of Eq. (\ref{Eqfinal1}), with the values of the model parameters taken from Table S1 in Ref.~\cite{kiv14_nature}. 
Note that the negative number given by the inference method in the high IPTG experiment signals that the actual value of ${\overline {\cal T}}_{E\to \mu}$ 
cannot be distinguished from zero, which is indeed the theoretical prediction. In contrast, the estimated and theoretical results for 
${\overline {\cal T}}_{\mu\to E}$ do not agree, as the inference method yields finite values in all cases whereas the theoretical values diverge. 
\begin{table}[!h]
  \centering
  \renewcommand{\arraystretch}{1.5}
  \begin{tabular}{|c|c|c|c|}
    \hline
    {\bf Conc. of IPTG} & Low & Intermediate & High \\
    \hline
    $\overline {\cal T}_{E\to\mu}$(in h$^{-1}$) (theo.) &$0.033$ & $0.034$& 0\\    
    $\overline {\cal T}_{E\to\mu}$ (simul.) & $0.031
    $& $0.034
    $& $-0.011 
    $\\
    \hline
    $\overline {\cal T}_{\mu\to E}$ (theo.) & $\infty$ & $\infty$ & $\infty$\\    
    $\overline {\cal T}_{\mu\to E}$ (simul.) & $0.202 
    $ & $0.123 
    $ & $0.347
    $\\
    \hline
  \end{tabular}
  \caption{Comparison between the theoretical values of the transfer entropy rates $\overline {\cal T}_{E\to\mu}$ and $\overline {\cal T}_{\mu\to E}$ for the model 
of Ref.~\cite{kiv14_nature} and the values inferred from simulation data. 
Averages are taken over $100$ times series of duration  $10^6$ min, sampled every $1$ min.}
\label{table OU}
\end{table} 
\begin{figure}[!h]
  \centering
  {\includegraphics[scale=0.5]{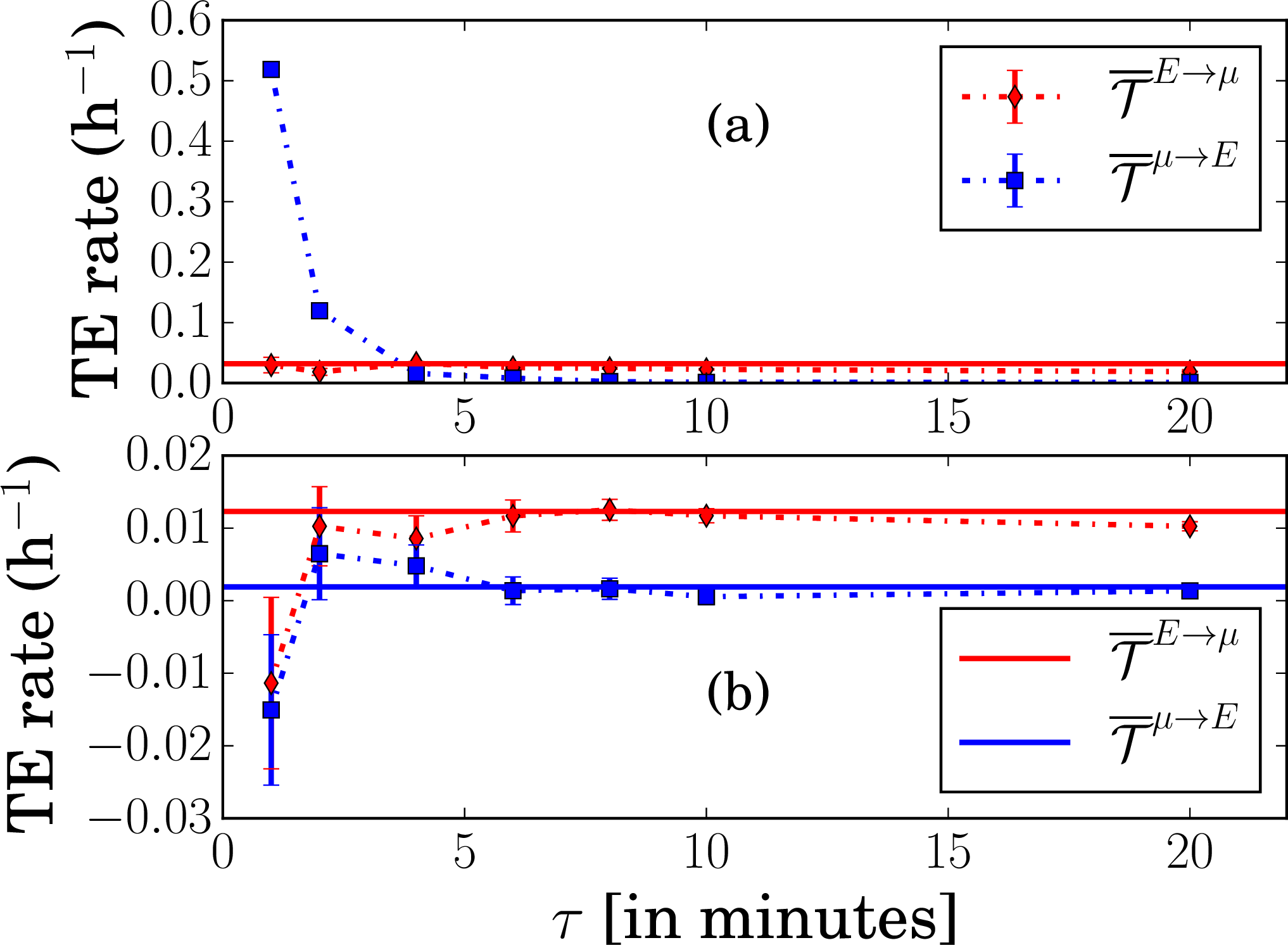}}
  \caption{Transfer entropy rates ${\overline {\cal T}}_{E \to \mu}$ and ${\overline {\cal T}}_{\mu \to E}$ in the low IPTG experiment: 
(a) Original model of Ref.~\cite{kiv14_nature} (b) Modified model where $N_E$ is a white noise. The symbols are the estimates from the
 inference method when varying the sampling time $\tau$, and the solid lines are the theoretical predictions from Eq. (\ref{Eqfinal1}) in (a) 
and from Eqs. (\ref{Eqfinal2}) in (b). Note that ${\overline {\cal T}}_{\mu \to E}$ diverges as $\tau$ goes to zero in (a) but not (b).
\label{Fig3}
}
\end{figure}

This behavior is due to the absence of a white noise source directly affecting the dynamical evolution of $x$ in the set of Eqs. (\ref{Eqxuvy}). Indeed, as pointed out in Ref.~\cite{sch00_prl} and also observed above in Fig \ref{fig:te_a_jidt}, a TE rate diverges when the coupling between the variables is deterministic.  In the model of Ref.~\cite{kiv14_nature}, this feature can be traced back to the fact that the noise $N_E$ affecting the enzyme concentration  is colored with a finite relaxation time $\beta_E^{-1}$. Therefore, when taking the limit  $\tau \to 0$ in Eq. (\ref{TE rate}), one explores a time interval $\tau<\beta_E^{-1}$ where $N_E$ is not really random.  This is illustrated in Fig \ref{Fig3}a that corresponds to the low IPTG experiment: we see that the estimate of ${\overline {\cal T}}_{\mu \to E}$ with the inference method is indeed diverging when the sampling time $\tau$ approaches zero. On the other hand,  as expected, ${\overline {\cal T}}_{E \to \mu}$ remains finite and the points nicely lie on the plateau determined by Eq. (\ref{Eqfinal1}).


The obvious and simplest way to cure this undesirable feature of the original model is to treat $N_E$ as a purely white noise, which amounts to taking the limit $\beta_E^{-1} \to 0$. In fact, it is noticeable that the values of $\beta_E^{-1}$ extracted from the fit of the correlation functions in Ref.~\cite{kiv14_nature} (resp. $\beta_E^{-1}=10.7,9.9$ and $8.15$ min for the low, intermediate, and high IPTG concentrations) are significantly smaller than the time steps $\tau_{exp}$ used for collecting the data  (resp. $\tau_{exp}=28, 20$ and $15.8$ min). Therefore, it is clear that the experimental data are not precise enough to decide whether $N_E$ is colored or not. 
This issue does not arise for the other relaxation times in the model, $\beta_{\mu}^{-1}=\beta_G^{-1}$ and $\mu_0^{-1}$, which are much longer (at least for the low and intermediate IPTG concentrations), and can be  correctly extracted from the experimental data.

\begin{figure*}[hbt]
\includegraphics[scale=0.5]{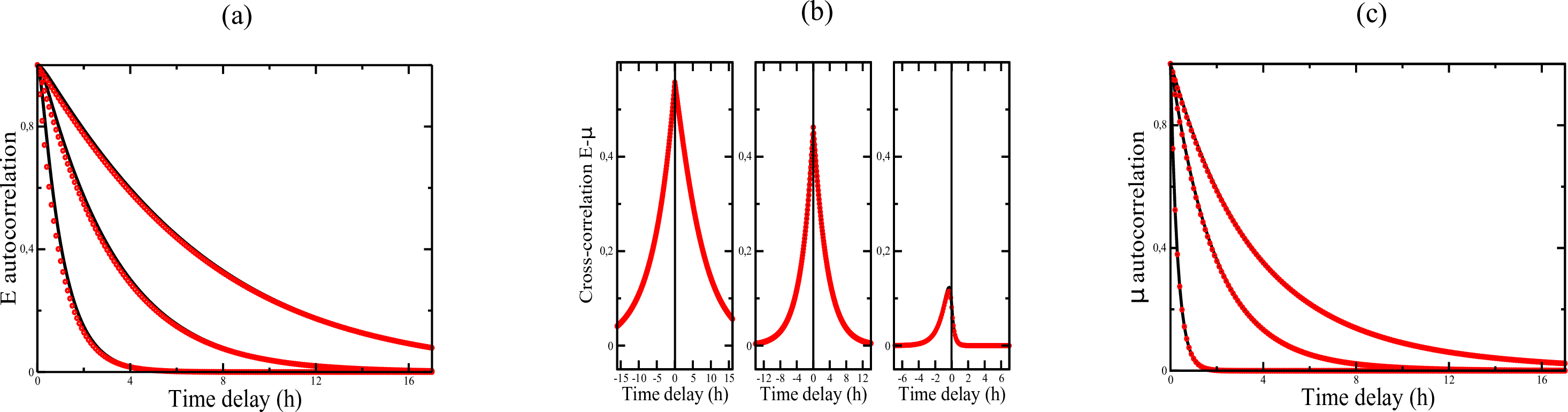}
\caption{\label{Fig2} (a) Autocorrelation function $R_{\mu\mu}(\tau)$ for the three IPTG concentrations. 
Black lines: original model of Ref.~\cite{kiv14_nature}, red circles: simplified model where $N_E$ is a white noise. 
(b) Same as (a) for $R_{EE}(\tau)$. (c) Same as (a) for $R_{E\mu}(\tau)$
\label{Fig:autocorr}
}
\end{figure*}

We thus propose to modify the model of Ref.~\cite{kiv14_nature} by describing $N_E$ as a Gaussian white noise with variance 
$\langle N_E(t)N_E(t')\rangle=2D_E\delta(t-t')$ and the same intensity as the colored noise in the original model, i.e. $D_E=\eta_E^2/\beta_E$ 
(which yields $D_E\approx 0.188 h,0.100h,0.031h$ for the three IPTG concentrations). 
Unsurprisingly, this modification does not affect the auto and cross-correlation functions used to fit the data, as shown in Fig \ref{Fig:autocorr} (see also section on Methods for a detailed calculation). 
On the other hand, the values of  ${\overline {\cal T}}_{E\to \mu}$ are changed (compare Tables \ref{table OU} and \ref{table3}) and, 
more importantly, ${\overline {\cal T}}_{\mu\to E}$, given by Eq. (\ref{Eqfinal2}) is now finite. As a result, the model predicts that the difference
 $\Delta {\overline {\cal T}}_{E\to\mu}={\overline {\cal T}}_{E\to \mu}-{\overline {\cal T}}_{\mu\to E}$  is positive at low and intermediate
 IPTG concentrations and becomes negative at high concentration, which is in agreement with the direct analysis of the experimental data in 
Table \ref{table:unconcat}. In contrast, $\Delta {\overline { \cal T}}_{E\to\mu}$ was always negative in the original model as ${\overline {\cal T}}_{\mu\to E}$ 
is infinite. 
\begin{table}[!h]
  \centering
  \renewcommand{\arraystretch}{1.2}
  \begin{tabular}{|c|c|c|c|}
    \hline
    {\bf Conc. of IPTG} & Low & Intermediate & High \\
    \hline
    $\overline {\cal T}_{E\to\mu}$ (h$^{-1}$)  & $1.23\cdot 10^{-2}$ & $8.2\cdot 10^{-3}$& $0$\\
    \hline
    $\overline  {\cal T}_{\mu\to E}$ (h$^{-1}$) & $1.9\cdot 10^{-3}$& $5\cdot 10^{-4}$& $2.97\cdot 10^{-2}$\\
    \hline  
  $\Delta \overline  {\cal T}_{E\to\mu}$ (h$^{-1}$)  & $1.04\cdot 10^{-2}$& $7.7\cdot 10^{-3}$& $-2.97\cdot 10^{-2}$ \\
  \hline
  \end{tabular}
  \caption{Theoretical values of the transfer entropy rates $\overline {\cal T}_{E\to\mu}$ and $\overline {\cal T}_{\mu\to E}$ and their 
difference in the modified model.}
\label{table3}
\end{table}

This new behavior of the TE rates is also manifest when the inference method is applied to the time series generated by the model and the sampling time $\tau$ is varied. As observed in Fig \ref{Fig3}b, the inferred value of ${\overline {\cal T}}_{\mu\to E}$ no longer diverges as $\tau \to 0$ (compare the vertical scale with that in Fig \ref{Fig3}a). The estimates of ${\overline {\cal T}}_{E\to \mu}$ and ${\overline {\cal T}}_{\mu\to E}$  are also in good agreement with the theoretical predictions, except for the shortest value of $\tau$ which is equal to the time step $\Delta t=1$ min used to numerically integrate the equations. It worth mentioning, however, that the error bars increase as $\tau$ is decreased. 


While the change in the sign of $\Delta {\overline {\cal T}}_{E\to\mu}$ is now confirmed by the model, 
which is the main outcome of our analysis, one may also wonder whether the numerical values  in Table \ref{table:unconcat} are recovered. This requires to multiply the rates in Table \ref{table3} by the experimental sampling times $\tau_{exp}$ which are different in each experiment, as indicated above. One then observes significant discrepancies for the low and intermediate IPTG experiments. 
We believe that the problem arises from the presence of many short time series in the set of experimental data. This is a important issue that needs to be examined 
in more detail since it may be  difficult to obtain long time series in practice.
\begin{figure}[!h]
  \centering
  {\includegraphics[width=7cm]{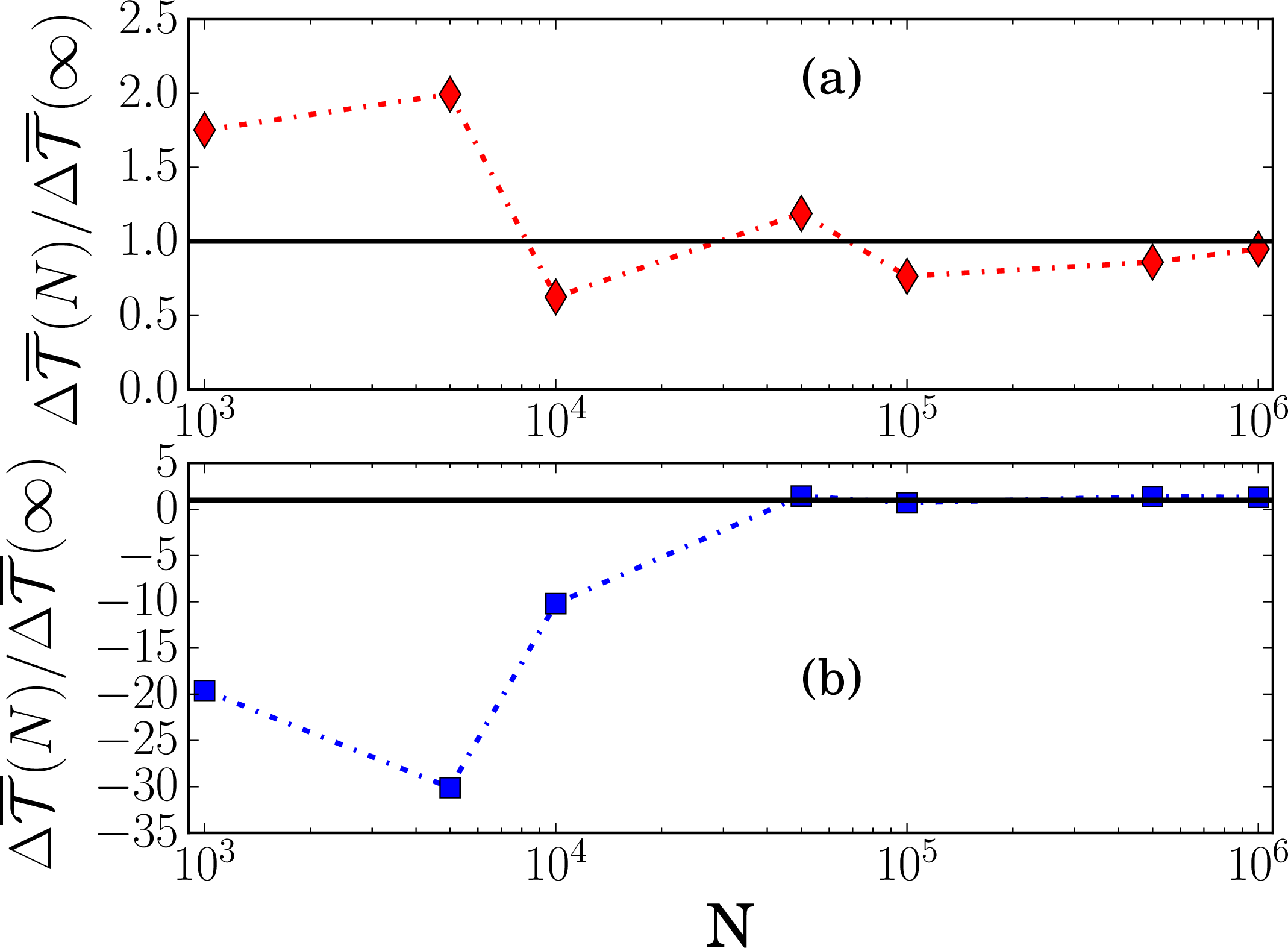}}
  \caption{Inferred values of $\Delta {\cal \overline T}_{E\to\mu}$ for the low IPTG experiment as a function of the length $N$ of the time series 
generated by the modified model. Panels (a) and (b) correspond to  sampling times $\tau=6$ min and $\tau=1$ min, respectively. 
$\Delta {\cal \overline T}_{E\to\mu}(\infty)$ is the exact asymptotic value. 
  \label{Fig4}
  }
\end{figure}

 
To this aim, we have studied the convergence of the estimates of $\Delta {\cal \overline T}_{E\to\mu}$  to the exact asymptotic value as a function of $N$, the length of the time series generated by the model in the stationary regime.  As shown in Fig \ref{Fig4}, the convergence with $N$ is slow, which means that one can make significant errors in the estimation of  $\Delta {\cal \overline T}_{E\to\mu}$ if $N$ is small. On the other hand, the convergence can be greatly facilitated by choosing a 
value of the sampling time which is not too short (but of course shorter than the equilibration time of the system),  
for instance $\tau=6$min instead of $1$ min in the case considered in Fig \ref{Fig4}. The important observation is that the sign of 
$\Delta {\cal \overline T}_{E\to\mu}$ is then correctly inferred even with  $N\approx1000$. In contrast, with $\tau=1$min, this is only 
possible for much longer series, typically $N\approx 50000$. This is an encouraging indication for experimental studies, as the overall 
acquisition time of the data can be significantly reduced. 



Finally, we briefly comment on the results for the information flows ${\cal I}^{flow}_{E\to \mu}$ and ${\cal I}^{flow}_{\mu\to E}$. 
As already pointed out, the fact that  the noises acting on the two random variables are correlated invalidates  inequality (\ref{ineq Iflow}). 
This is indeed what is observed in Table \ref{TableS2}. 
It is also noticeable that  ${\cal I}^{flow}_{E\to \mu}\ne-{\cal I}^{flow}_{\mu\to E}$, except in
 the high IPTG experiment where $T_{\mu E}=0$. 
\begin{table*}[!h]
  \centering
  \renewcommand{\arraystretch}{1.5}
  \begin{tabular}{|c|c|c|c|}
    \hline
    {\bf Conc. of IPTG} & Low & Intermediate & High \\
    \hline
    ${\overline {\cal T}}^{E\to\mu}$, analytical &$ 0.0123$ & $0.0082$& 0\\
    ${\overline {\cal T}}^{E\to\mu}$, simulation & $0.0128 \pm 6\cdot 10^{-4}$& $0.0064  \pm 6\cdot 10^{-4}$ & $ -0.0002 \pm 5 \cdot 10^{-4}$\\
    \hline
    $\overline {\cal T}^{\mu\to E}$, analytical & $0.0019 $ & $0.0005$ & $0.0297$\\
 $\overline {\cal T}^{\mu\to E}$, simulation & $0.0023 \pm 6 \cdot 10^{-4}$ & $ 0.0012 \pm 6 \cdot 10^{-4}$ & $0.0215 \pm 7 \cdot 10^{-4}$ \\   
    \hline
    ${\cal I}^{flow}_{E\to\mu}$, analytical & $0.0751$ & $0.092$ & $-0.0214$\\
${\cal I}^{flow}_{E\to\mu}$, simulation & $0.076 \pm 10^{-3}$ & $0.09 \pm 8 \cdot 10^{-4}$ & $-0.018 \pm 8 \cdot 10^{-4}$  \\  
    \hline
    ${\cal I}^{flow}_{\mu \to E}$, analytical & $0.0455$ & $0.0743$ & $0.0214$\\
    ${\cal I}^{flow}_{\mu \to E}$, simulation & $0.047 \pm 10^{-3}$ & $0.072 \pm 10^{-3}$ & $0.015 \pm 10^{-3}$\\
    \hline
  \end{tabular}
  \caption{Comparison between the theoretical values of the TE rates and the information flows for the modified model 
  and the values inferred from simulation data (all quantities are expressed in h$^{-1}$). The analysis was performed with a sampling $\tau=6$ min and $100$  time series of $10^6$ points.}
\label{TableS2}
\end{table*}

\section*{Discussion and conclusion}

A challenge when studying any biochemical network is to properly identify the 
direction of information.
In this work, using the notion of transfer entropy, we have characterized the directed flow of information 
between the single cell growth rate and the gene expression, 
using a method that goes beyond what could be obtained from correlation functions, 
or from other inference techniques which do not exploit dynamical information. 

Another crucial challenge in the field is to properly model the various noise components.
It turns out that biological systems are generally non-bipartite due the presence of an extrinsic component in the noise.
The present work provides on the one hand analytical expressions for the magnitude of the transfer entropy (or at least an upper bound on it)
and of the information flow when the system is not bipartite, and, on the other hand a numerical method to infer the TE in all cases. 
Furthermore, we have shown that one can correctly infer the sign of the TE difference even with short time series 
by properly choosing the sampling time (see Ref. \cite{Barnett2017} for more details on the dependence of TE on the sampling time).



To conclude, we would like to emphasize that the transfer entropy is a general tool to 
identify variables which are relevant for time series prediction \cite{Tishby2000}.
As such, the method has a lot of potential beyond the particular application covered in this paper:
Predicting the current or future state of the environment by sensing it is an adaptation strategy followed by biological systems
which can be understood using information-theoretic concepts \cite{Tostevin2009,Hartich2016}.
Similarly, during evolution, biological systems accumulate information from their environment, process 
it and use it quasi-optimally to increase their own fitness \cite{Kobayashi2015,Halabi2009}.
In this context, transfer entropy-based methods have the potential to 
identify the directional interactions in co-evolution processes, which could be 
for instance the genomic evolution of a virus compared to that of its antigenes \cite{Smith2004}.
With the recent advances in high-throughput techniques and 
experimental evolution, 
we might soon be able to predict reliably the evolution of biological 
systems \cite{Laessig2017}, and without doubt  tools of information theory will play a key role 
in these advances.

\section*{Methods}

In this section, we provide a detailed analysis of the information-theoretic quantities 
for the various models considered in this paper. The section is organized as follows:

\begin{itemize}
\item
Basic information-theoretic measures
\item
Transfer entropy and information flow in the feedback cooling model
\item
Transfer entropy rates and information flows in the model of Ref.~\cite{kiv14_nature} for a metabolic network
\item
Transfer entropy rates and information flows in the modified model for the metabolic network
\end{itemize}


\subsection*{Basic information-theoretic measures}
\label{app:A}
Below we briefly recall some definitions and properties of the information-theoretic measures.
A fundamental quantity is the Shannon entropy which quantifies the uncertainty associated with the measurement $x$ of a random variable $X$:
\be
H(X)=-\sum_{x } P(x) \ln P(x),
\ee
where $P(x)$ is the probability that event $x$ is realized, given an ensemble of possible outcomes. With this convention, the entropy is measured in nats. Similarly, for two random variables $X$ and $Y$, one defines the joint Shannon entropy
\be
H(X,Y)=-\sum_{x,y} P(x,y) \ln P(x,y),
\ee
and the conditional Shannon entropy 
\be
H(X | Y)=-\sum_{x,y}  P(x,y) \ln P(x | y)\ ,
\ee
where $P(x,y)$ and $P(x\vert y)$ are joint and conditional probability distribution functions, respectively. The mutual  information $I(X:Y)$ is then a {\it symmetric} measure defined as
\begin{align}
I(X:Y) &= \sum_{x,y} P(x,y) \ln \frac{P(x,y)}{P(x)P(y)},  \nn \\
       &= H(X) - H(X | Y) \nn \\
       &= H(Y)-H(Y\vert X)\ ,
\end{align}
which quantifies the reduction of the uncertainty about $X$ (resp. $Y$) resulting from the knowledge of the value of $Y$ (resp$X$). The more strongly $X$ and $Y$ are correlated, the larger $I(X:Y)$ is.

These notions can be readily extended to random processes $X= \{ X_i\}$ and $Y= \{ Y_i\}$ viewed as collections of individual random variables sorted by an integer time index $i$. The mutual information between the ordered time series  $ \{ x_i\}$ and $\{ y_i\}$, realizations of $X$ and $Y$, is then defined as 
\be
 I(X: Y) = I(Y:X) \equiv \sum_{ \{ x_i, y_i \} }  P(x_i,y_i) \ln\frac{P(x_{i}, y_i)}{P(x_{i})P(y_i)}\ ,
\label{def mutual_info}
\ee
 and characterizes the {\it undirected} information exchanged between the two processes.
The conditional mutual information is defined similarly.

In contrast, the transfer entropy $T_{X \to Y}$ is a information-theoretic measure that is both {\it asymmetric} and {\it dynamic} as it captures  the amount of information that a source process $X$ provides about the next state of a target process $Y$. More precisely, as defined by Eq. (1) in the introduction,  
\begin{align}
 T_{X \to Y}=\sum_i \: [ I(Y_{i+1}:\pmb{X}_i^{(l)},\pmb{Y}_i^{(k)})-I(Y_{i+1}:\pmb{Y}_i^{(k)})],
\label{def TE-bis}
\end{align}
where $k$ and $l$ define the lengths of the process histories, i.e., $\pmb{Y}_i^{(k)}=\{Y_{i-k+1},\cdots, Y_i\}$ and $\pmb{X}_i^{(l)}=\{X_{i-l+1},\cdots, X_i\}$. In this work, we have focused on a history length of $1$ (i.e. $k=l=1$) and denoted the corresponding TE by $\overline T_{X \to Y}$.  Hence, 
 ${\overline T}_{X \to Y}=\sum_i [ H(Y_{i+1}|Y_i)-H(Y_{i+1}|X_i,Y_i)]$, which is an upper bound to $T_{X \to Y}(k,l)$ for $l=1$ 
 when the joint process $\{ X,Y \}$ obeys a Markovian dynamics~\cite{Hartich2016}. 

On the other hand, the information flow from $X$ to $Y$ is defined as the time-shifted mutual information 
\begin{align}
{\cal I}^{flow}_{X \to Y}=\sum_i [I(Y_i : X_i) - I (Y_i:X_{i+1})], 
\label{def info_flow}
\end{align}
and informs on the reduction of uncertainty in $Y_i$ when knowing about $X_{i+1}$ as compared to what we had with $X_i$ only. In practice, ${\cal I}^{flow}_{X \to Y}$ can be obtained by shifting in time one time series with respect to the other one.
Contrary to the transfer entropy which is always a positive quantity, the information flow ${\cal I}^{flow}_{X \to Y}$ may be  negative or positive, depending on whether $X$ sends information to $Y$ (or $X$ gains control of $Y$), or $Y$ sends information to $X$ (or $X$ looses control over $Y$ ).
In a bipartite system one has ${\cal I}^{flow}_{X \to Y} = - {\cal I}^{flow}_{Y \to X}$ in the stationary regime. This is no longer true when the system is non-bipartite.



\subsection*{Transfer entropy and information flow in the feedback cooling model}

We first recall the theoretical expressions of the transfer entropy rates and the information flows for the feedback-cooling model 
described by Eqs. (\ref{Lang}). These quantities were computed  in Ref.~\cite{hor14_njp}. The transfer entropy rates in the stationary state are given by
\begin{align}
  {\cal T}_{V \to Y} &= \frac{\gamma}{2m}\left(\sqrt{1+\frac{2T}{\gamma\sigma^2}}-1\right)  \nn \\
  {\cal T}_{Y \to V} &= \frac{1}{2\tau_r}\left(\sqrt{1+\frac{a^2\sigma^2}{2\gamma T}}-1\right).
\label{TEfeedback}
\end{align}
Note that $2T/(\gamma\sigma^2)$ is the signal-to-noise ratio that quantifies the relative size of the measurement accuracy to the thermal diffusion of the velocity.  Accordingly, the TE rate $ {\cal T}_{V \to Y}$ diverges when the control is deterministic. The information flow ${\cal I}^{flow}_{V \to Y}$ is given by
\be
{\cal I}^{flow}_{V \to Y}= \frac{\gamma}{m} \left( \frac{T \langle y^2\rangle}{m \vert {\bf \Sigma}\vert} - 1 \right)
\label{I_flow}
\ee 
where $\vert {\bf \Sigma}\vert$ is the determinant of the covariance matrix. The analytical expressions of the elements of the matrix, $\langle v^2\rangle,\langle y^2\rangle$ and $\langle v y\rangle$, are given by Eqs. (A2) in Ref.~\cite{hor14_njp}. In contrast with $ {\cal T}_{V \to Y}$, the information flow  ${\cal I}^{flow}_{V \to Y}$ remains finite as the noise intensity vanishes.

The upper bounds to the transfer entropies (see Eq. (2)) were computed in Ref.~\cite{ito14_nc} in the general case of coupled linear Langevin equations. For the feedback cooling model, one obtains 
\begin{align}
 {\overline {\cal T}}_{V \to Y} &= \frac{1}{2\sigma^2\langle y^2\rangle}\vert {\bf \Sigma}\vert \nn \\
  {\overline {\cal T}}_{Y \to V} &= \frac{a^2}{4\gamma k_BT\langle v^2\rangle}\vert {\bf \Sigma}\vert\ .
\label{TEbar}
\end{align}

\begin{figure}[!h]
\centering
{\includegraphics[width=7.5cm]{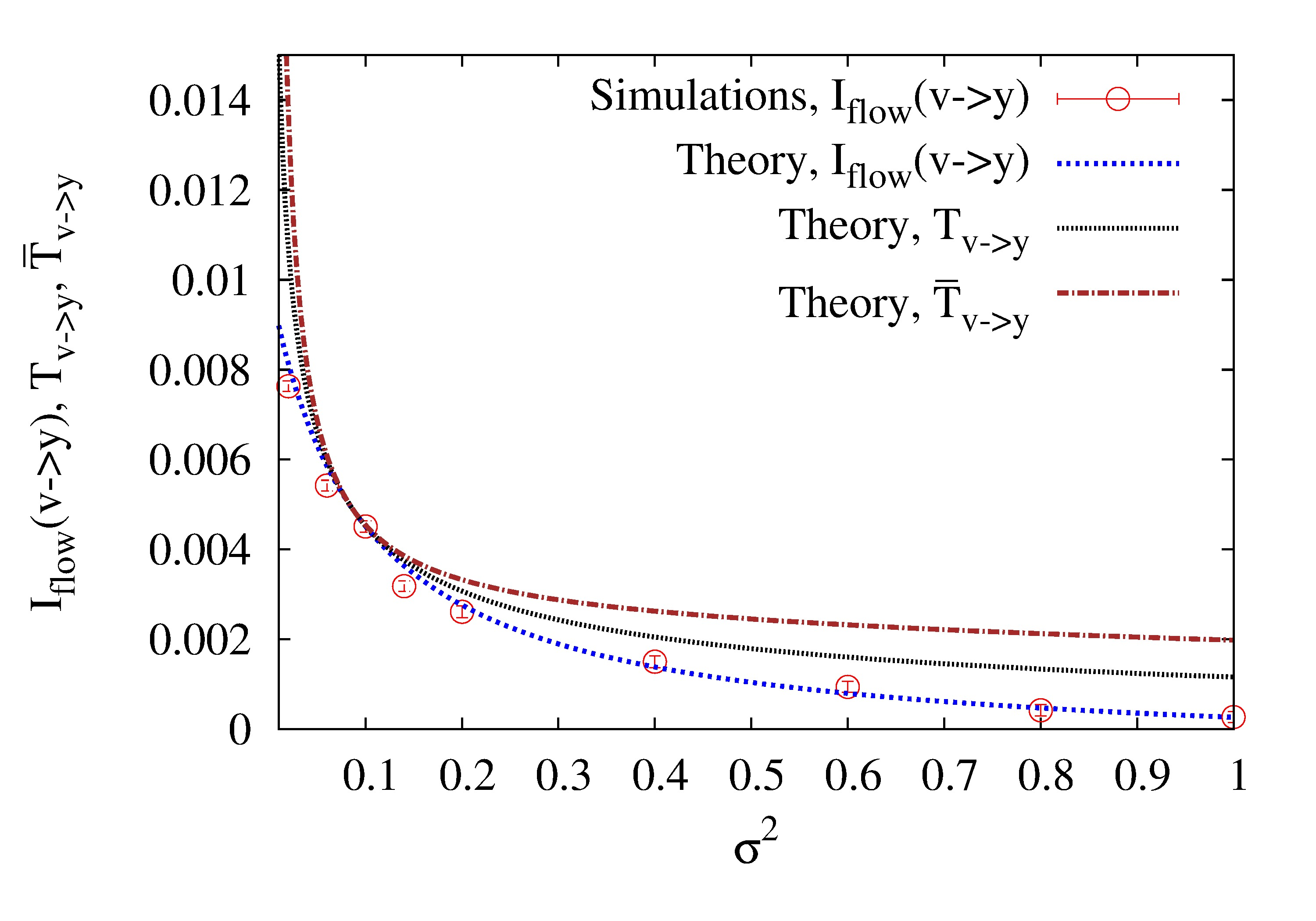}}
 \caption{$ {\cal T}_{V \to Y}, {\overline {\cal T}}_{V \to Y}$ and ${\cal I}^{flow}_{V \to Y}$ as a function of the noise intensity $\sigma^2$. The parameters of the model are $T = 5,\gamma = m = 1, \tau_r =0.1$ and $a=-0.7$.
\label{fig:I_flow} 
}
\end{figure}

As shown in Fig \ref{fig:te_a_jidt}, the estimate of the transfer entropy obtained by the inference method is in good agreement with the theoretical value (we stress that the figure shows the rates multiplied by the sampling time $\tau=10^{-3}$).  
In Fig \ref{fig:I_flow}, we also obtain satisfactory agreement between inferred value of the information flow ${\cal I}^{flow}_{V \to Y}$ 
and theoretical value, when representing these quantities against the noise intensity $\sigma^2$. These results of this figure 
confirm the inequalities ${\cal I}^{flow}_{V \to Y}\le  {\cal T}_{V \to Y}\le  {\overline {\cal T}}_{V \to Y}$.


\subsection*{Transfer entropy rates and information flows in the model of Ref.~\cite{kiv14_nature} for a metabolic network }
\label{app:B}

\subsubsection*{Stationary distributions and correlation functions}

We first compute the stationary probability distributions (pdfs) associated with Eqs. (\ref{Eqxuvy}) were the coefficients $a_j$ and $b_j$ are given by 
\begin{align} 
a_1&=-[\mu_E +\mu_0 T_{\mu E}(T_{EG}-1)]\nonumber\\
a_2&=\mu_0\nonumber\\
a_3&=-\mu_0 T_{EG}\nonumber\\
a_4&=\mu_0(T_{EG}-1)\nonumber\\
b_1&=T_{\mu E}[\beta_G-\mu_E -\mu_0 T_{\mu E}(T_{EG}-1)]\nonumber\\
b_2&=\mu_0T_{\mu E}\nonumber\\
b_3&=\beta_G-\beta_{\mu}-\mu_0 T_{\mu E}T_{EG}\nonumber\\
b_4&=\mu_0 T_{\mu E}(T_{EG}-1)-\beta_G \ .
\label{coeff a_i}
\end{align}
We recall that $\mu_E=\mu_0(1+T_{\mu E}-T_{EE})$ sets the timescale of $E$-fluctuations~\cite{kiv14_nature}. Since Eqs. \eqref{Eqxuvy} 
describe a set of coupled Markovian Ornstein-Uhlenbeck  processes, the  stationary pdf $p_{xuvy}(x,u,v,y)$ is Gaussian and given by
\begin{align} 
\label{Eqpxuvy}
p_{xuvy}(x,u,v,y)=\frac{1}{(2\pi)^2\sqrt{\vert {\bf \Sigma} \vert}} e^{-\frac{1}{2}(x,u,v,y).{\bf \Sigma}^{-1}.(x,u,v,y)^T}\ ,
\end{align}
where ${\bf \Sigma}$ is the covariance matrix which obeys the Lyapunov equation~\cite{Risken1989}
\begin{align} 
\label{EqLya}
{\bf A}{\bf \Sigma}+{\bf \Sigma}{\bf A}^T=2{\bf D}\ ,
\end{align}
where 
\[
{\bf A}=\left(
\begin{array}{cccc}
   -a_1&-a_2 & -a_3&-a_4 \\
  0&\beta_E &0 &0    \\
  0&0 &\beta_{\mu}& 0   \\
  -b_1&-b_2 &-b_3 &-b_4
\end{array}
\right), \, \, \, \rm{and } \, \, \,
{\bf D}=\left(
\begin{array}{cccc}
   0&0& 0&0 \\
  0&\beta_E\eta_E^2  &0 &0    \\
  0&0 &\beta_{\mu}\eta_{\mu}^2&\beta_{\mu}\eta_{\mu}^2    \\
  0&0 & \beta_{\mu}\eta_{\mu}^2& \beta_G\eta_G^2 + \beta_{\mu}\eta_{\mu}^2
\end{array}
\right) \ .
\]
 
The solution of Eq. (\ref{EqLya}) reads
\begin{align} 
\label{Eqsigma}
\sigma_{11}&=\frac{\mu_0^2}{\mu_E}\Big[\frac{\eta_E^2}{\mu_E+\beta_E}+\frac{\eta_{\mu}^2}{\mu_E+\beta_{\mu}}+\frac{(T_{EG}-1)^2}{\mu_E+\beta_G}\eta_G^2\Big]\nonumber\\
\sigma_{12}&=\sigma_{21}=\frac{\mu_0}{\mu_E+\beta_E}\eta_E^2\nonumber\\
\sigma_{13}&=\sigma_{31}= \frac{-\mu_0}{\mu_E+\beta_{\mu}}\eta_{\mu}^2\nonumber\\
\sigma_{14}&=\sigma_{41}=\frac{\mu_0}{\mu_E}\Big[\frac{\mu_0T_{\mu E}}{\mu_E+\beta_E}\eta_E^2+\frac{(\mu_0T_{\mu E}-\mu_E)}{\mu_E+\beta_{\mu}}\eta_{\mu}^2 \nonumber \\
&+\frac{(T_{EG}-1)\big[\mu_0T_{\mu E}(T_{EG}-1)+\mu_E\big]}{\mu_E+\beta_G}\eta_G^2\Big] \nonumber\\
\sigma_{22}&=\eta_E^2\nonumber\\
\sigma_{23}&=0\nonumber\\
\sigma_{24}&=\sigma_{42}=\frac{\mu_0T_{\mu E}}{ \mu_E+\beta_E} \eta_E^2\nonumber\\
\sigma_{33}&=\eta_{\mu}^2\nonumber\\
\sigma_{34}&=\sigma_{43}=\frac{\mu_E+\beta_{\mu}-\mu_0T_{\mu E}}{\mu_E+\beta_{\mu}}\eta_{\mu}^2\nonumber\\
\sigma_{44}&=\frac{\mu_0^2T_{\mu E}^2}{ \mu_E(\mu_E+\beta_E)}\eta_E^2+\frac{\big[(\mu_0T_{\mu E}-\mu_E)^2+ \mu_E\beta_{\mu}\big]}{ \mu_E(\mu_E+\beta_{\mu})}\eta_{\mu}^2 \nonumber \\
&+\frac{\mu_0^2T_{\mu E}^2(T_{EG}-1)^2+\mu_E\big[\mu_E+\beta_G\big]}{\mu_E(\mu_E+\beta_{G})}\eta_G^2 \nonumber  \\
&+\frac{2\mu_0T_{\mu E}(T_{EG}-1)\big]}{\mu_E(\mu_E+\beta_{G})} \eta_G^2
\end{align}

\begin{figure}[hbt]
\begin{center}
\includegraphics[width=8cm]{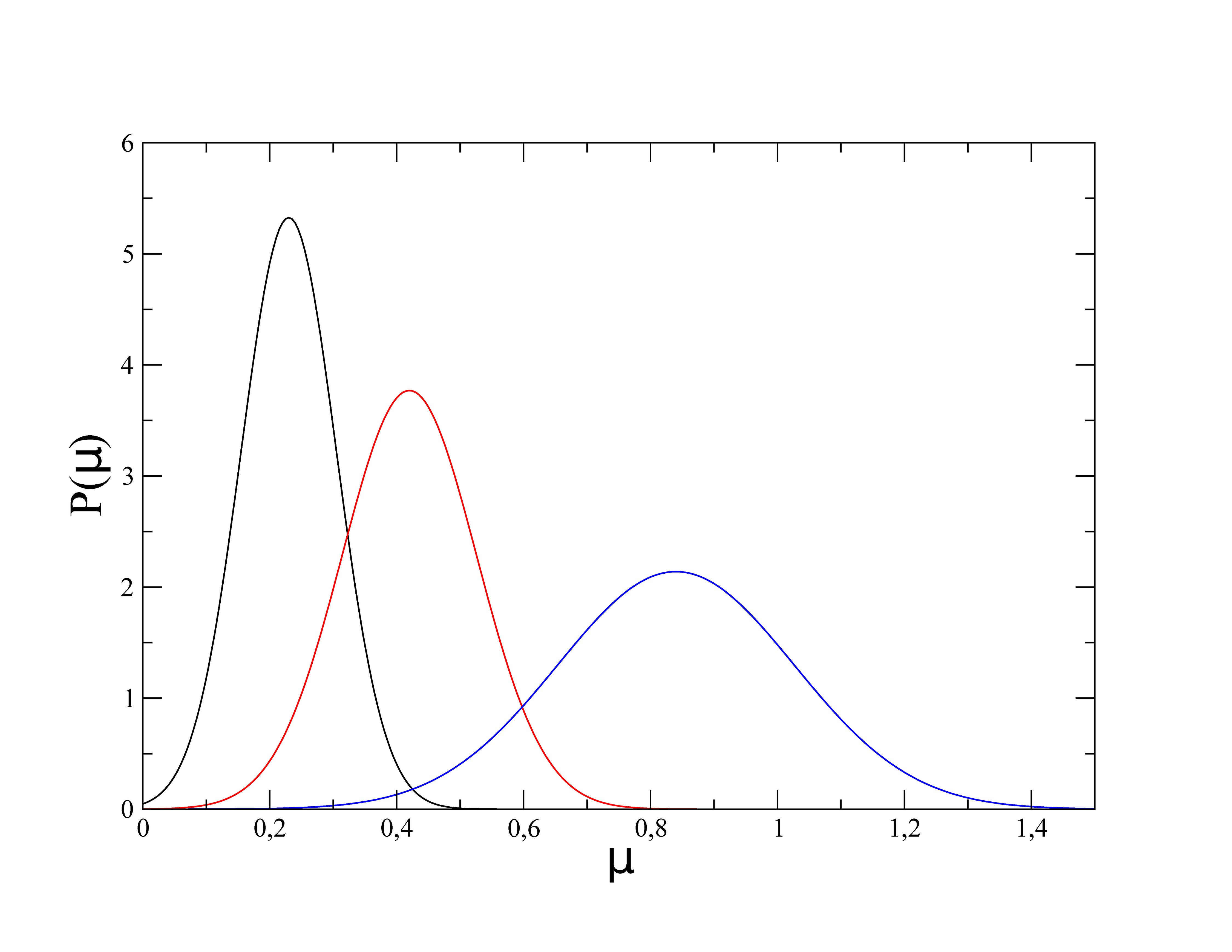}
\caption{\label{Fig1} Steady-state probability distribution of the growth rate for the three IPTG concentrations: 
 low (black), intermediate (red), high (blue).}
\end{center}
\end{figure}

From this we can compute all  marginal pdfs, in particular
\begin{align} 
p_{xy}(x,y)&=\frac{1}{2\pi\sqrt{\sigma_{11}\sigma_{44}-\sigma_{14}^2}}e^{-\frac{1}{2}\frac{\sigma_{44}x^2-2\sigma_{14}xy+\sigma_{11}y^2}{\sigma_{11}\sigma_{44}-\sigma_{14}^2}}\ ,
\end{align}
and
\begin{align} 
p_{x}(x)&=\frac{1}{\sqrt{2\pi \sigma_{11}}}e^{-\frac{x^2}{2\sigma_{11}}}\nonumber\\ 
p_{y}(y)&=\frac{1}{\sqrt{2\pi \sigma_{44}}}e^{-\frac{y^2}{2\sigma_{44}}}\ .
\end{align}
As an illustration, the steady-state pdf $p(\mu)=\frac{1}{\mu_0}p_y(y=\frac{\mu-\mu_0}{\mu_0})$ is plotted in Fig \ref{Fig1} for the three 
different IPTG concentrations (low, intermediate, and high). The agreement with the experimental curves displayed in Fig 1d of Ref.~\cite{kiv14_nature} is satisfactory.

For completeness, we also quote the expressions of $R_{pp}(0)$ and $R_{p\mu}(0)$ (properly normalized) obtained from the definition
 $\delta p/(\mu_0 E_0)=\delta \dot E/(\mu_0 E_0)+\delta \mu/\mu_0+\delta E/E_0=(T_{EE}-T_{EG}T_{\mu E})x+u-T_{EG}(v-y)$:
\begin{align}
R_{pp}(0)&=(T_{EE}-T_{EG}T_{\mu E})^2\sigma_{11}+\sigma_{22}+T_{EG}^2(\sigma_{33}+\sigma_{44}) \nonumber\\
&+2(T_{EE}-T_{EG}T_{\mu E})[\sigma_{12}+T_{EG}(\sigma_{14}-\sigma_{13})] \nonumber \\
&+2T_{EG}\sigma_{24}-2T_{EG}^2\sigma_{34} \\
R_{p\mu}(0)&=\frac{(T_{EE}-T_{EG}T_{\mu E})\sigma_{14}+\sigma_{24}+T_{EG}(\sigma_{44}-\sigma_{34})}{\sqrt{R_{pp}(0)R_{\mu\mu}(0)}}
\end{align}
with $R_{\mu \mu}(0)=\sigma_{44}$.

The correlation functions $R_{\mu \mu}(\tau)$, $R_{EE}(\tau)$, and $R_{E\mu}(\tau)$, obtained by taking the inverse Fourier transform of Eqs. (6) in the Supplementary Information of \cite{kiv14_nature} are plotted in Fig \ref{Fig2}.  In passing, we correct a few misprints in  these equations: 
i) The correct expression of  $R_{\mu \mu}(\tau)$ is obtained by replacing $A_E(\tau)$ by $R_{EE}(\tau)$  in the first term of Eq. (12) in the Supplementary Information of \cite{kiv14_nature}.
ii) Eq.~\ref{postulated} corresponds to $R_{E \mu}(\tau)$ and {\it not} to $R_{\mu E}(\tau)=R_{E \mu}(-\tau)$. Eq. (8) then gives the correct expression of $R_{E\mu}(\tau)$ (and not of $R_{\mu E}(\tau)$) provided the function $A_X(\tau)$ defined in Eq. (10) is altered.  For $\tau \ge 0$, one should have 
\begin{align} 
A_X(\tau)=\theta_X^2\frac{\mu_0}{2\beta_X(\beta_X+\mu_E)}e^{-\beta_X t}\ .
\end{align}

\subsubsection*{Transfer entropy rates}

We now address the computation of the conditional probabilities $p_{x' y'}^y(y,t+\tau\vert x',y',t)$ and $p_{y'}^y( y,t+\tau\vert y',t)$ at first order in $\tau$. This will allow us to obtain the expressions of the upper bounds to the transfer entropy rates defined by
\begin{align} 
{\overline {\cal T}}_{X\to Y}&=\lim_{\tau \to 0}\frac{I[y_{t+\tau}:x_t,y_t]-I[y_{t+\tau}:y_{t}]}{\tau}\nonumber\\
{\overline {\cal T}}_{Y\to X}&=\lim_{\tau \to 0}\frac{I[x_{t+\tau}:x_t,y_t]-I[x_{t+\tau}:x_{t}]}{\tau}\ ,
\end{align}
where $I$ is the mutual information, for instance $I[y_{t+\tau}:x_t,y_t]=\int dy\:dx'\:dy'\:p_{x'y'}^{y}(y,t+\tau;x',y',t)\ln [p_{x'y'}^{y}(y,t+\tau;x',y',t)/[p_{y}(y)p_{xy}(x',y')]$ in the steady state (where  $p_{xy}(x',y')$ and $p_y(y)$ become time independent pdfs). Therefore,
\begin{align} 
\label{EqT}
{\overline {\cal T}}_{X\to Y}&=\lim_{\tau \to 0}\frac{1}{\tau} \int dy\:dx'\:dy'\:p_{x'y'}^{y}(y,t+\tau;x',y',t)\nonumber\\\
&\times \ln \frac{p_{x'y'}^{y}(y,t+\tau\vert x',y',t)}{p_{y'}^y(y,t+\tau\vert y',t)}\nonumber\\
{\overline {\cal T}}_{Y\to X}&=\lim_{\tau \to 0}\frac{1}{\tau}\int dy\:dx'\:dy'\:p_{x'y'}^{x}(x,t+\tau; x',y',t) \nonumber\\
&\times \ln \frac{p_{x'y'}^{x}(x,t+\tau\vert x',y',t)}{p_{x'}^x(x,t+\tau\vert x',t)}\ .
\end{align}
Note that the actual transfer entropy rates are defined as 
\begin{align} 
{\cal T}_{X\to Y}&=\lim_{\tau \to 0}\frac{I[y_{t+\tau}:x_t,\{y_{t'}\}_{t'\le t}]-I[y_{t+\tau}:\{y_{t'}\}_{t'\le t}]}{\tau} \nonumber\\
 {\cal T}_{Y\to X}&=\lim_{\tau \to 0}\frac{I[x_{t+\tau}:\{x_{t'}\}_{t'\le t},y_t]-I[x_{t+\tau}:\{x_{t'}\}_{t'\le t}]}{\tau} \ .
\end{align}
where $\{x_{t'}\}_{t'\le t}$ and $\{y_{t'}\}_{t'\le t}$ denote the full trajectories of $x_t$ and $y_t$ in the time interval $[0,t]$. Since the  present model is not bipartite, the calculation of these quantities is a nontrivial task that is left aside. 


The two-time distributions $p_{x'y'}^{y}(y,t+\tau;x',y',t)$ and  $p_{x'y'}^{x}(x,t+\tau;x',y',t)$ are given by
\begin{align}
\label{Eqpx1y1y}
p_{x'y'}^{y}(y,t+\tau;x',y',t)&= \int dx\:du\:dv\:du'\:dv'\:p_{{\bf z}'}^{{\bf z}}({\bf z},t+\tau\vert {\bf z}',t)p_{xuvy}({\bf z}')\nonumber\\
p_{x'y'}^{x}(x,t+\tau;x',y',t)&= \int dy\:du\:dv\:du'\:dv'\:p_{{\bf z}'}^{{\bf z}}({\bf z},t+\tau\vert {\bf z}',t)p_{xuvy}({\bf z}')
\end{align}
where $p_{{\bf z}'}^{{\bf z}}({\bf z},t+\tau\vert {\bf z}',t)$ is the transition probability from the state ${\bf z}'=(x',u',v',y')$ at time $t$ to the state ${\bf z}=(x,u,v,y)$ at time $t+\tau$. From the definition of the Fokker-Planck operator ${\cal L}_{FP}$ associated with the 
$4$-dimensional diffusion process described by Eqs. \ref{Eqxuvy},
the transition probability for small times is given by~\cite{Risken1989}
\begin{align}
&p_{{\bf z}'}^{{\bf z}}({\bf z},t+\tau\vert {\bf z}',t)=[1+\tau {\cal L}_{FP}({\bf z},t)+{\cal O}(\tau^2)]\delta({\bf z}-{\bf z}')\nonumber\\
&=\delta({\bf z}-{\bf z}')-\tau \sum_{i=1}^4 \partial_{z_i}\big[g_i({\bf z}')-\sum_j\frac{\theta_{i,j}^2}{2}\partial_{z_j}\big]\delta({\bf z}-{\bf z}') 
\end{align}
where $g_i({\bf z})$ is the drift coefficient in the equation for $z_i$ (with $z_1=x,z_2=u,z_3=v,z_4=y$), $\theta_{2,2}=\theta_{E},\theta_{3,3}=\theta_{3,4}=\theta_{\mu},\theta_{4,4}=\sqrt{\theta_{\mu}^2+\theta_G^2}$ and all other $\theta_{i,j}$ being equal to $0$. 
 
Let us first consider the calculation of ${\overline {\cal T}}_{X\to Y}$. By integrating $p_{{\bf z}'}^{{\bf z}}({\bf z},t+\tau\vert {\bf z}',t)$ over $x$, $u$, and $v$,  we readily obtain
\begin{align*} 
p_{{\bf z}'}^{y}(y,t+\tau\vert {\bf z}',t)=\delta(y-y')-\tau \partial_{y}\big[g_y({\bf z}')-\beta_{\mu}\eta_{\mu}^2\partial_{v}-(\beta_{\mu}\eta_{\mu}^2+\beta_G \eta_G^2)\partial_y\big]\delta(y-y')+{\cal O}(\tau^2)
\end{align*}
where the terms involving $\partial_x,\partial_u,\partial_v$ cancel due to natural boundary conditions. Hence,
\begin{align} 
p_{{\bf z}'}^y(y &,t+\tau; {\bf z}',t)=p_{{\bf z}'}^{y}(y,t+\tau\vert {\bf z}',t)p_{xuvy}({\bf z}')\nonumber\\
&=\delta(y-y')p({\bf z}')-\tau p_{xuvy}({\bf z}') \times \nonumber \\
& \partial_{y} \big[g_y({\bf z}')-\beta_{\mu}\eta_{\mu}^2\partial_{v}-(\beta_{\mu}\eta_{\mu}^2+\beta_G \eta_G^2)\partial_y\big]\delta(y-y') ,
\end{align}
which yields 
\begin{align} 
\label{Eqpxyy0}
p_{x'y'}^y(&y,t+\tau; x',y',t)=\delta(y-y')p_{xy}(x',y') -\tau p_{xy}(x',y')\partial_{y}\big[\bar g_y(x',y') \nonumber \\
  &-(\beta_{\mu}\eta_{\mu}^2+\beta_G \eta_G^2)\partial_y\big]\delta(y-y'). 
\end{align}
after integration over $u'$ and $v'$, where we have defined the averaged drift coefficient
\begin{align} 
\bar g_{y}(x,y)=\frac{1}{p_{xy}(x,y)}\int du\:dv\:g_y({\bf z})p_{xuvy}({\bf z})\ .
\label{g bar}
\end{align}
We thus finally obtain
\begin{align} 
\label{Eqpxyy1}
p_{x'y'}^y(y,&t+\tau\vert x',y',t)=\delta(y-y')-\tau \partial_{y}\big[\bar g_y(x',y') \nonumber \\
&-(\beta_{\mu}\eta_{\mu}^2+\beta_G \eta_G^2)\partial_y\big]\delta(y-y')+{\cal O}(\tau^2)\ .
\end{align}
Similarly, by also integrating $p_{{\bf z}'}^y(y,t+\tau; x',y',t)$ over $x'$, we obtain
\begin{align} 
\label{Eqpyy}
p_{y'}^y(y,t &+\tau\vert y',t)=\delta(y-y')-\tau \partial_{y}\big[{\bar {\bar g}}_y(y')-(\beta_{\mu}\eta_{\mu}^2 \nonumber \\
&+\beta_G \eta_G^2)\partial_y\big] \delta(y-y')+{\cal O}(\tau^2)\ .
\end{align}
where
\begin{align} 
{\bar {\bar g}}_{y}(y)&=\frac{1}{p_{y}(y)}\int dx\:du\:dv\: g_y({\bf z})p_{xuvy}({\bf z})\nonumber\\
&=\frac{1}{p_{y}(y)}\int dx\:{\bar g}_y(x,y)p_{xy}(x,y)\ .
\label{g bar bar}
\end{align}
Due to the linearity of Eqs.~\eqref{Eqxuvy} and the Gaussian character of the pfds, one simply has $\bar g_{y}(x,y)=ax+by$ and ${\bar {\bar g}}_{y}(y)=cy$, where $a,b,c$ are complicated functions of the model parameters which we do not display here. 

Eq. (\ref{Eqpxyy1}) (resp. Eq. (\ref{Eqpyy})) merely shows that $p_{x'y'}^y(y,t+\tau\vert x',y',t)$ (resp. $p_{y'}^y(y,t+\tau\vert y',t)$) at the lowest order in $\tau$ is identical to the transition probability associated with an Ornstein-Uhlenbeck process with drift coefficient $\bar g_{y}(x,y)$ (resp. ${\bar {\bar g}}_{y}(y)$) and diffusion coefficient $\beta_{\mu}\eta_{\mu}^2+\beta_G \eta_G^2$. To proceed further, it is then convenient to use to the Fourier integral representation of the $\delta$ function and re-express  $p_{x'y'}^y(y,t+\tau\vert x',y',t)$  and $p_{y'}^y(y,t+\tau\vert y',t)$ for small times as
\begin{align} 
p_{x'y'}^y(y,t+\tau\vert x',y',t)=\frac{1}{2\sqrt{\pi(\beta_{\mu}\eta_{\mu}^2+\beta_G \eta_G^2)\tau}}e^{-\frac{1}{4(\beta_{\mu}\eta_{\mu}^2+\beta_G \eta_G^2)\tau}[y-y'- \tau\bar g_{y}(x',y')]^2}
\end{align}
and
\begin{align} 
p_{y'}^y(y,t+\tau\vert y',t)=\frac{1}{2\sqrt{\pi(\beta_{\mu}\eta_{\mu}^2+\beta_G \eta_G^2)\tau}}e^{-\frac{1}{4(\beta_{\mu}\eta_{\mu}^2+\beta_G \eta_G^2)\tau}[y-y'- \tau{\bar {\bar g}}_{y}(y')]^2} \ .
\end{align}
up to corrections of the order $\tau^2$~\cite{Risken1989}. This leads to 
\begin{align} 
\ln \frac{p_{x'y'}^y(y,t+\tau\vert x',y',t)}{p_{y'}^y(y,t+\tau\vert y',t)}&=\frac{1}{4(\beta_{\mu}\eta_{\mu}^2+\beta_G \eta_G^2)}\big[2(y-y')-\tau[\bar g_{y}(x',y')+{\bar {\bar g}}_{y}(y')]\big] \nn \\
& \times \big[\bar g_{y}(x',y')-{\bar {\bar g}}_{y}(y')\big]\ ,
\end{align}
and from Eq. (\ref{Eqpxyy0}) and the definition of the transfer entropy rate [Eq. (\ref{EqT})],
\begin{align} 
4(\beta_{\mu}\eta_{\mu}^2+\beta_G \eta_G^2){\overline {\cal T}}_{X\to Y}&=\lim_{\tau \to 0} \frac{1}{\tau}\int dy\:dx'\:dy' p_{x'y'}^y(y,t+\tau;x',y',t)\big[2(y-y') \nn \\
&-\tau[\bar g_{y}(x',y')+{\bar {\bar g}}_{y}(y')]\big]\big[\bar g_{y}(x',y')-{\bar {\bar g}}_{y}(y')\big]\nonumber\\
&=\lim_{\tau \to 0} \frac{1}{\tau}\int dy\:dx'\:dy' p_{xy}(x',y')\Big[\delta(y-y')-\tau \partial_{y}[\bar g_y(x',y') \nn \\ 
&-(\beta_{\mu}\eta_{\mu}^2+\beta_G \eta_G^2)\partial_y]\delta(y-y')\Big]\nonumber\\
&\times \big[2(y-y')-\tau[\bar g_{y}(x',y')+{\bar {\bar g}}_{y}(y')]\big]\big[\bar g_{y}(x',y')-{\bar {\bar g}}_{y}(y')\big]
\end{align}
We then use
\begin{align} 
\int dy\: (y-y')\Big[\delta(y-y')-\tau \partial_{y}[\bar g_y(x',y')-(\beta_{\mu}\eta_{\mu}^2+\beta_G \eta_G^2)\partial_y]\delta(y-y')\Big]=\tau \bar g_y(x',y')\ ,
\end{align}
and
\begin{align} 
\int dx'\:p_{xy}(x',y')\bar g_y(x',y')&=p_y(y'){\bar {\bar g}}_{y}(y')=\int dx' \: p_{xy}(x',y'){\bar {\bar g}}_{y}(y')\ ,
\end{align}
to finally arrive at Eq. \eqref{Eqfinal1}, namely
\begin{align} 
\label{Eqfinal1-bis}
{\overline {\cal T}}_{X\to Y}&=\frac{1}{4(\beta_{\mu}\eta_{\mu}^2+\beta_G \eta_G^2)}\int dx\:dy\: p_{xy}(x,y)\big[\bar g_{y}^2(x,y)-{\bar {\bar g}}_{y}^2(y)\big]\ .
\end{align}
A similar expression can be found in Ref.~\cite{Hartich2016} (see Eq. (A.31) in that reference). Note also that  the result given in Ref.~\cite{ito14_nc} is obtained as a special case.

Inserting into Eq. (\ref{Eqfinal1}) the values of the parameters given in Table S1 of Ref.~\cite{kiv14_nature}, we obtain the values given in Table 2. Note that ${\overline {\cal T}}_{E\to \mu}=0$ for the high IPTG concentration because $T_{\mu E}=0$, and therefore $\mu(t)$ no longer depends on $E(t)$ as can be seen from Eq. \eqref{postulated}.

There is no need to detail the calculation of ${\overline {\cal T}}_{\mu\to E}$ (i.e. ${\overline {\cal T}}_{Y\to X}$) because it goes along the same line, with $y$ replaced by $x$.
The crucial difference is that there is no white noise acting on $\dot x$. Therefore, the denominator in Eq. (\ref{Eqfinal1}), which is the variance of the noise $\xi_y$, is replaced by $0$. This implies  that ${\overline {\cal T}}_{\mu\to E}$ is infinite. 

\subsubsection*{Information flows}

The information flows ${\cal I}^{flow}_{X\to Y}$ and ${\cal I}^{flow}_{Y\to X}$ are derived from the time-shifted mutual informations $I[x_{t+\tau}:y_t]$ and $I[y_{t+\tau}:x_t]$. Specifically, 
\begin{align} 
\label{EqIflow}
{\cal I}^{flow}_{X\to Y}&=\lim_{\tau \to 0} \frac{I[x_t:y_t]-I[x_{t+\tau}:y_t]}{\tau}\nonumber\\
{\cal I}^{flow}_{Y\to X}&=\lim_{\tau \to 0} \frac{I[y_t:x_t]-I[y_{t+\tau}:x_t]}{\tau}\ .
\end{align} 

Let us first consider the second flow ${\cal I}^{flow}_{Y\to X}$  which requires the knowledge of $p_{x'}^y(y,t+\tau;x', t)$ whose expression is  obtained by integrating Eq. (\ref{Eqpxyy0}) over $x'$. This yields 
\begin{align} 
p_{x'}^y(y,t+\tau; x',t)=p_{xy}(x',y)-\tau \partial_{y}\big[\bar g_y(x',y) \nonumber \\ 
-(\beta_{\mu} \eta_{\mu}^2+\beta_G \eta_G^2)\partial_y \big]p_{xy}(x',y)+{\cal O}(\tau^2)\ .
\end{align}
Hence
\begin{align} 
  I[y_{t+\tau}:x_t]&=\int dx'\:dy\: p_{x'}^y(y,t+\tau; x',t)\nonumber\\ 
  &\hspace{1cm}\times\ln \frac{p_{x'}^y(y,t+\tau; x',t)}{p_y(y)p_x(x')}\nonumber\\
&=I[y_t:x_t]-\tau \int dx\:dy\: \partial_{y}\big[\bar g_y(x,y) \nonumber \\
&-(\beta_{\mu}\eta_{\mu}^2+\beta_G \eta_G^2)\partial_y \big]p_{xy}(x,y)\ln \frac{p_{xy}(x,y)}{p_y(y)p_x(x)}.
\end{align} 
We finally obtain
\begin{align} 
&{\cal I}^{flow}_{Y\to X}=\int dx\:dy\: \partial_{y}\big[\bar g_y(x,y)p_{xy}(x,y) \nonumber \\
&-(\beta_{\mu}\eta_{\mu}^2+\beta_G \eta_G^2)\partial_yp_{xy}(x,y)\big]\ln \frac{p_{xy}(x,y)}{p_y(y)p_x(x)}\ .
\end{align} 
A similar calculation yields
\begin{align} 
{\cal I}^{flow}_{X\to Y}&=\int dx\:dy\: \partial_{x}\big[\bar g_x(x,y)p_{xy}(x,y)\big]\ln \frac{p_{xy}(x,y)}{p_y(y)p_x(x)}\ ,
\end{align} 
where 
\begin{align} 
\bar g_{x}(x,y)=\frac{1}{p_{xy}(x,y)}\int du\:dv\:g_x({\bf z})p_{xuvy}({\bf z})
\end{align}
is an  averaged drift coefficient. Contrary to the case of the transfer entropy rate ${\overline {\cal T}}_{Y\to X}$, the absence of a white noise acting on $\dot x$ does not lead to an infinite result for  ${\cal I}^{flow}_{Y\to X}$. In fact, one has the symmetry relation
\begin{align}
{\cal I}^{flow}_{X\to Y}=-{\cal I}^{flow}_{Y\to X}\ ,
\end{align}
which is readily obtained by noting that $p_{xy}(x,y)$, the stationary solution of the Fokker-Planck equation, satisfies the equation
\begin{align} 
\partial_x[& \bar g_{x}(x,y)p_{xy}(x,y)]+\partial_y[\bar g_{y}(x,y)p_{xy}(x,y)] \nonumber \\
&-(\beta_{\mu}\eta_{\mu}^2+\beta_G \eta_G^2)\frac{\partial^2 }{\partial y^2}p_{xy}(x,y)=0 \ .
\end{align}
Inserting  the numerical values of the parameters given in Table S1 of Ref.~\cite{kiv14_nature}, we obtain the values given in Table \ref{TableS1} below. Interestingly,  $ {\cal I}^{flow}_{E \to \mu}$ decreases as the IPTG concentration increases and that it  becomes negative at high concentration.
\begin{table}[!h]
  \centering
  \renewcommand{\arraystretch}{1.5}
  \begin{tabular}{|c|c|c|c|}
    \hline
    {\bf Conc. of IPTG} & Low & Intermediate & High \\
    \hline
   $ {\cal I}^{flow}_{E \to \mu}$(in h$^{-1}$)&$0.0148$ & $0.0088$& -0.0243\\    
    \hline
  \end{tabular}
  \caption{Theoretical values of ${\cal I}^{flow}_{X\to Y}=-{\cal I}^{flow}_{Y\to X}$ in the original model of Ref.~\cite{kiv14_nature}}
\label{TableS1}
\end{table}

\subsection*{Transfer entropy rates and information flows in the modified model for the metabolic network}

We now repeat the above calculations for the modified model where $N_E$ is treated as a white noise. Eliminating again the variable $w$ (i.e. $N_G$) in favor of $y$, the new set of equations that describe the stochastic dynamics and replace 
Eqs.~\ref{Eqxuvy} reads
\begin{align}
\label{Eqxuvynew}
\dot x&=-\big[\mu_E +\mu_0 T_{\mu E}(T_{EG}-1)\big]x-\mu_0 T_{EG}v \nonumber\\
&+\mu_0(T_{EG}-1)y+\xi_x \nonumber \\
\dot v&=-\beta_{\mu} v+\xi_{\mu}\nonumber\\
\dot y&=T_{\mu E}\big[\beta_G-\mu_E -\mu_0 T_{\mu E}(T_{EG}-1)\big]x+\big[\beta_G-\beta_{\mu} \nonumber \\
&-\mu_0 T_{\mu E}T_{EG}\big]v+\big[\mu_0 T_{\mu E}(T_{EG}-1)-\beta_G\big]y+\widetilde\xi_{y}\ ,
\end{align}
where we have defined the  white noises $\xi_x=\mu_0N_E$ and $\widetilde\xi_{y}=\xi_y+T_{\mu E}\xi_{x}$ satisfying $\langle \xi_x(t)\xi_x(t')\rangle=2D_E\mu_0^2\delta(t-t')$ and $\langle\widetilde\xi_{y}(t)\widetilde\xi_y(t')\rangle=(\theta_{\mu}^2+\theta_G^2+2D_E\mu_0^2T_{\mu E}^2)\delta(t-t')$, respectively. These two noises are  correlated, with $\langle \xi_x(t)\widetilde\xi_y(t')\rangle=2D_E\mu_0^2T_{\mu E}\delta(t-t')$.

The pdfs and the correlation functions can be computed as before. In fact, it is clear that this simply amounts to taking the limit $\beta_E \to \infty$ with $D_E=\eta_E^2/\beta_E$ finite in the previous equations (for instance in Eqs. (\ref{Eqsigma}) for the covariances). The new correlation functions are plotted in Fig \ref{Fig2}. As expected, they are almost indistinguishable from those obtained with the original model and they fit the experimental data just as well (this of course is also true for the pdfs).

Much more interesting are the results for the transfer entropy rates and the information flows. Again, there is no need to repeat the calculations as they follow the same lines as before. We now obtain
\begin{align} 
\label{Eqfinal2}
{\overline {\cal T}}_{X\to Y}&=\frac{1}{4(\beta_{\mu}\eta_{\mu}^2+\beta_G \eta_G^2+D_E\mu_0^2T_{\mu E}^2)} \times \nonumber \\
&\int dx\:dy\: p_{xy}(x,y)\big[\bar g_{y}^2(x,y)-{\bar {\bar g}}_{y}^2(y)\big]\\
{\overline {\cal T}}_{Y\to X}&=\frac{1}{4D_E\mu_0^2}\int dx\:dy\: p_{xy}(x,y)\big[\bar g_{x}^2(x,y)-{\bar {\bar g}}_{x}^2(x)\big]\ ,
\end{align}
where 
\begin{align} 
\bar g_{x}(x,y)&=\frac{1}{p_{xy}(x,y)}\int du\:g_x(x,v,y)p_{xvy}(x,v,y)\\
\bar g_{y}(x,y)&=\frac{1}{p_{xy}(x,y)}\int du\:g_y(x,v,y)p_{xvy}(x,v,y)\ ,
\end{align}
and 
\begin{align} 
{\bar {\bar g}}_{x}(x)&=\frac{1}{p_{x}(x)}\int dy\:{\bar g}_x(x,y)p_{xy}(x,y)\\
{\bar {\bar g}}_{y}(y)&=\frac{1}{p_{y}(y)}\int dx\:{\bar g}_y(x,y)p_{xy}(x,y)\ .
\end{align}
(Again, $g_x(x,v,y)$ and $g_y(x,v,y)$ denote the drift coefficients in Eqs.(\ref{Eqxuvynew})). The crucial difference with the results for the original model is that ${\overline {\cal T}}_{Y\to X}$ is now finite. 
Similarly, we have
\begin{align} 
\dot I^{flow}_{X\to Y}&=\int dx\:dy\: \partial_{x}\big[\bar g_x(x,y)p_{xy}(x,y) \nonumber\\
&-D_E\mu_0^2\partial_xp_{xy}(x,y)\big]\ln \frac{p_{xy}(x,y)}{p_y(y)p_x(x)}\\
\dot I^{flow}_{Y\to X}&=\int dx\:dy\: \partial_{y}\big[\bar g_y(x,y)p_{xy}(x,y) \nonumber\\
&-(\beta_{\mu}\eta_{\mu}^2+\beta_G \eta_G^2+D_E\mu_0^2T_{\mu E}^2)\partial_yp_{xy}(x,y)\big]  \nonumber\\
&\times \ln \frac{p_{xy}(x,y)}{p_y(y)p_x(x)}\ .
\end{align}

The numerical values of ${\overline {\cal T}}_{E\to \mu}$ and ${\overline {\cal T}}_{\mu \to E}$ are given in Table \ref{table3}. For completeness, we also compare these values with the estimates obtained by the inference method  in Table \ref{TableS2}. We see that satisfactory results are obtained by properly choosing the sampling time $\tau$. This is also true for the information flows ${\cal I}^{flow}_{E\to\mu}$ and ${\cal I}^{flow}_{\mu\to E}$.  It is worth noting that the symmetry relation $\dot I^{flow}_{E\to \mu}=-\dot I^{flow}_{\mu\to E}$ no longer holds, except for the high  IPTG concentration (as $T_{\mu E}=0$). This contrasts with the preceding case where $N_E$ was modeled by an Ornstein-Uhlenbeck noise.  We also observe that the information flows are not always smaller than the transfer entropy rates, contrary to what occurs in bipartite systems. Therefore, the concept of  a "sensory capacity" as introduced in Ref.~\cite{Hartich2016} is here ineffective.

\section*{Acknowledgments}
We acknowledge J. Lizier for many insightful comments regarding the numerical evaluation of transfer entropies, and L. Peliti for stimulating discussions.
S.L. thanks the Institute of Complex Systems (ISC-PIF), the Region Ile-de-France, and the Labex
CelTisPhyBio (No. ANR-10- LBX-0038) part of the IDEX PSL (No. ANR-10-IDEX-0001-02 PSL) for financial support.

\nolinenumbers

%
%
%


\end{document}